\documentclass[aps,superscriptaddress,preprint,nofootinbib,eqsecnum]{revtex4}
\usepackage{amsmath,amssymb,bm}
\usepackage{graphicx}
\usepackage{epstopdf}

\let\a=\alpha   \let\d=\partial

\let\w=\omega

\def\nn{\nonumber}

\let\pa=\partial
\newcommand{\bea}{\begin{eqnarray}}
\newcommand{\eea}{\end{eqnarray}}
\def\ba{\begin{array}}
\def\ea{\end{array}}
\def\del{\delta}

\def\ep{{\epsilon}}
\def\vep{{\varepsilon}}

\def\be{\begin{equation}}
\def\ee{\end{equation}}

\renewcommand\d{\partial}

\begin{document}
\vspace*{-1.5cm} \hfill {NSF-KITP-07-144} \vspace*{1cm}

\title{Theory of the Nernst effect near quantum phase transitions in condensed matter, and in dyonic black holes}

\author{Sean A. Hartnoll}\affiliation{Kavli Institute for Theoretical Physics,
University of California, Santa Barbara, CA 93106-4030, USA}

\author{Pavel K. Kovtun}\affiliation{Kavli Institute for Theoretical Physics,
University of California, Santa Barbara, CA 93106-4030, USA}

\author{Markus M\"uller}
\affiliation{Department of Physics, Harvard University, Cambridge MA
02138, USA}

\author{Subir Sachdev}
\affiliation{Department of Physics, Harvard University, Cambridge MA
02138, USA}

\begin{abstract}
\noindent
We present a general hydrodynamic theory of transport in the vicinity
of superfluid-insulator transitions in two spatial dimensions
described by ``Lorentz''-invariant quantum critical points. We allow for a weak impurity
scattering rate, a magnetic field $B$, and a deviation
in the density, $\rho$, from that of the insulator.
We show that the frequency-dependent
thermal and electric linear response functions,
including the Nernst coefficient, are fully determined by a single
transport coefficient (a universal electrical conductivity), the impurity scattering rate, and a few thermodynamic state variables.
With reasonable estimates for the parameters, our results
predict a magnetic field and temperature dependence of the Nernst signal which
resembles measurements in the cuprates,
including the overall magnitude.
Our theory predicts a ``hydrodynamic cyclotron mode'' which could
be observable in ultrapure samples. We also present exact results for the zero frequency
transport co-efficients
of a supersymmetric conformal field theory (CFT), which is solvable by the AdS/CFT correspondence.
This correspondence maps the $\rho$ and $B$ perturbations of the 2+1 dimensional CFT to electric and magnetic
charges of a black hole in the 3+1 dimensional anti-de Sitter space.
These exact results are found to be in full agreement
with the general predictions of our hydrodynamic analysis in the appropriate limiting regime.
The mapping of the hydrodynamic and AdS/CFT results under particle-vortex duality is also
described.
\end{abstract}

\date{June  2007}

\maketitle

\section{Introduction}
\label{sec:intro}

A key indication that the normal state of the cuprate
superconductors is aberrant came from the pioneering measurements
of the Nernst effect by Ong and collaborators
\cite{ong1,ong2,ong4,ong3}. Also of interest here are measurements of
the Nernst effect in Nb$_{0.15}$Si$_{0.85}$ films by
Behnia and collaborators \cite{behnia1,behnia2}. The Nernst co-efficient measures the
transverse voltage arising in response to an applied thermal gradient
in the presence of a magnetic field. The response of Fermi liquids
is associated with a weak particle-hole asymmetry in the spectrum
of the fermionic excitations near the Fermi level \cite{vadim}. The large
observed response, and its striking and unexpected dependence on
the magnetic field, temperature, and carrier concentration
indicated that an explanation starting from a metallic Fermi
liquid state could not be tenable.  Instead, Ong and collaborators
argued that their observations called for a description in terms
of a liquid of quantized vortices and anti-vortices in the
superconducting order (and its precursors) at low temperatures.

A complete theory of the dynamics of the vortex liquid state is so far lacking.
Ussishkin {\em et al.} \cite{ussishkin}
used a classical Gaussian theory of
superconducting fluctuations, and Mukerjee and Huse \cite{mukerjee} extended this
to a time-dependent Ginzburg-Landau model.
Podolsky {\em et al.} \cite{podolsky}
applied classical Langevin equations to a model of phase variables residing on the sites
of a hypothetical lattice. Anderson \cite{pwa} has taken a speculative view of the vortex liquid,
arguing against the conventional Debye-screening of vortex interactions.
While some experimental trends are successfully described by Refs.~\onlinecite{ussishkin,mukerjee,podolsky},
it would be useful to have a kinematic approach which
is also able to include quantum effects, and extends across the superfluid-insulator transition.
Quantum effects which surely play an more important role at lower temperatures, especially in the
underdoped
region. Indeed, it is the equal importance of thermal and quantum fluctuations which underlies
the difficulty in describing this vortex liquid.

As in the recent work by Bhaseen, Green, and Sondhi \cite{bgs},
this paper will advocate an approach departing from the quantum
critical region of a zero temperature ($T$) quantum phase
transition between a superconductor and an insulator. This is the
region where the primary perturbation from the physics of the
$T=0$ quantum critical point is the temperature. The single energy
scale, $k_B T$, then determines observable properties, including
the values of diffusion constants and transport co-efficients, in
a manner that has been discussed at length elsewhere
\cite{ssbook,m2cft}. The electrical conductivity of this quantum
critical system, which we denote $\sigma_Q$, will play a prominent
role in our results. In 2+1 dimensions, near quantum critical
points which obey hyperscaling properties, this conductivity is
given by
\cite{mpaf,wenzee,damle}
\begin{equation}
\sigma_Q = \frac{4e^2}{h} \Phi_\sigma, \label{ds}
\end{equation}
where in the quantum critical region $\Phi_\sigma$ is a universal dimensionless number dependent
only upon the universality class of the critical point.

\begin{figure}[t]
\centering \includegraphics[width=4in]{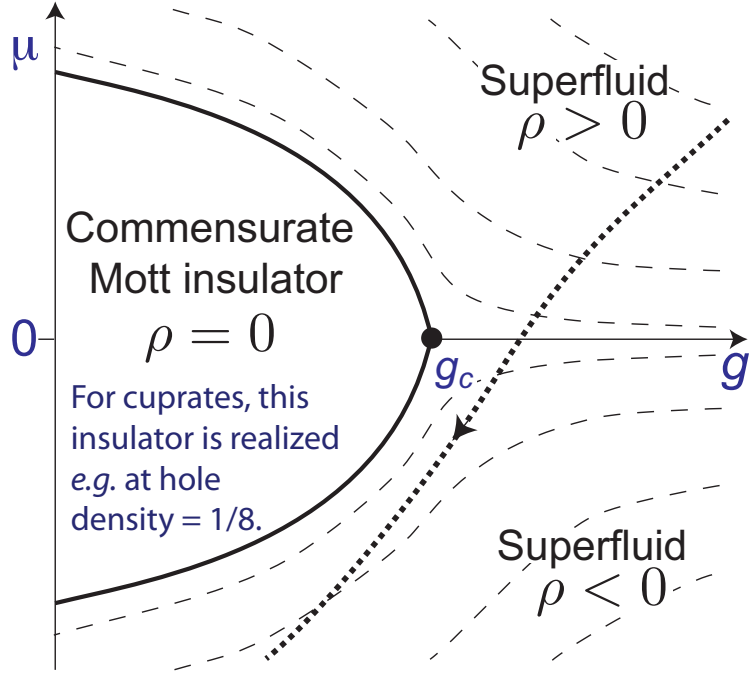}
\caption{Zero temperature ($T=0$), zero field ($B=0$) phase diagram in the vicinity of the quantum critical point
described by the CFT, represented by the filled circle. The coupling $g$ represents a parameter
which tunes between a superfluid and a Mott insulator which is at a density
commensurate with the underlying lattice. The chemical potential $\mu$ introduces variations
in the density and $\rho$ is difference in the density of pairs of holes in the superfluid from that in the Mott insulator.
The thin dashed lines are contours of constant $\rho$. In the application to the cuprate superconductors,
the Mott insulator with $\rho=0$ could be, {\em e.g.\/}, an insulating state at hole density $\delta_I=1/8$ in a generalized phase diagram; then $\rho = (\delta-\delta_I)/(2a^2)$, where $a$ is the lattice spacing.
The thick dotted line represents a possible trajectory of a particular compound as its
hole density is decreased; note that the ground state is always a superconductor
along this trajectory, even at $\delta=1/8$ (although there will be a dip in $T_c$ near $\delta=1/8$ as is also
clear from Fig.~\ref{phasediag2}). Note that the parent Mott insulator with zero hole density is not shown above. This paper will describe electrical and thermal transport in the above
phase diagram perturbed by an applied magnetic field $B$ and a small density of impurities.
}
\label{phasediag}
\end{figure}

\begin{figure}[t]
\centering \includegraphics[width=5in]{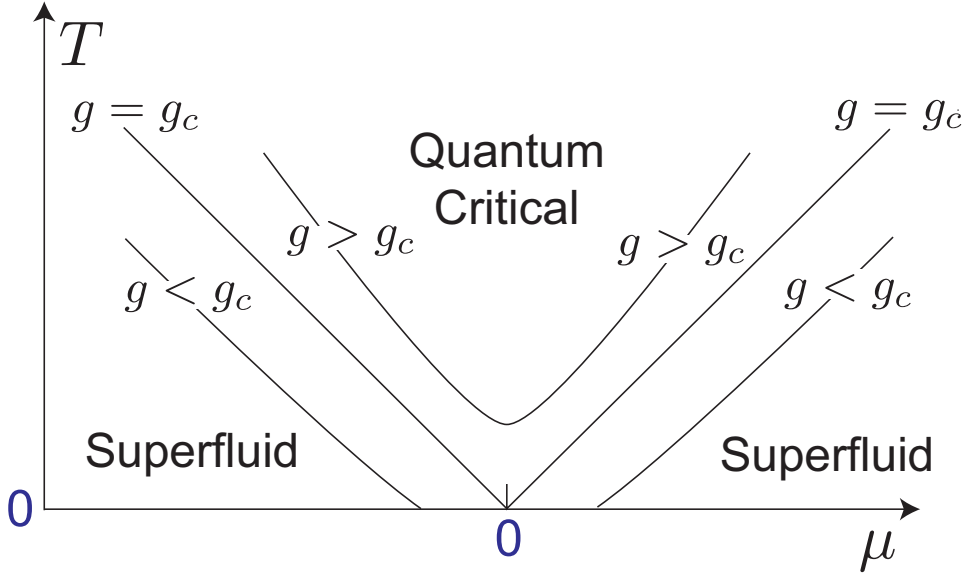}
\caption{Nonzero temperature ($T$) phase diagram at $B=0$ along three vertical cuts ({\em i.e.\/} fixed $g$) in Fig.~\ref{phasediag}. The lines indicate Kosterlitz-Thouless
phase transitions at $T=T_{KT}$ associated with the loss of superfluid order as a function
of $\mu$ for different values of $g$. At $g=g_c$, $T_{KT}/|\mu| $ is a universal number determined by the CFT at $g=g_c$, $\mu=0$ \cite{troyer}. This paper describes transport
properties in the non-superfluid region above $T_{KT}$, in the presence of an applied magnetic field $B$
and a small impurity scattering $\tau_{\rm imp}$. The results of the supersymmetric CFT solvable by AdS/CFT in Section~\ref{sec:dyon} are limited to $g=g_c$, but allow
arbitrary variations in $\mu$ and $B$ away from quantum criticality as long there is no phase transition into a superfluid (or other) state.
}
\label{phasediag2}
\end{figure}

The discussion so far applies, strictly speaking, only to systems which are exactly at the commensurate density for which a gapped Mott insulator can form. The cuprates, and other experimental systems, are not generically at these special densities, and so it is crucial to develop a theory that is applicable at generic densities. Such a theory will emerge as a special case of our more general results below. We allow the density to take values $\rho$ by applying a chemical potential $\mu$, as shown in Figs.~\ref{phasediag} and~\ref{phasediag2}.

We emphasize that $\rho$ measures the {\em deviation\/} in particle number density
from the density of the commensurate insulator \cite{balents1,lorenz}; so $\rho$ can be positive or negative,
and we will see that the sign of $\rho$ determines the sign of the Hall resistivity and other
transport coefficients. Also, purely as a {\em choice of convention\/}, we will measure $\rho$ in terms of density
of pairs of holes; this choice does not imply the degrees of freedom of the underlying
theory are Cooper pairs, although this is the case in the simplest model.
We will use general hydrodynamic arguments (specialized to
``relativistic'' quantum critical points) to show that the
frequency ($\omega$) dependent conductivity, $\sigma_{xx}$
at a generic density $\rho$ is given by
\begin{equation}
\sigma_{xx}  = \sigma_Q + \frac{4 e^2 \rho^2 v^2}{(\varepsilon+P)} \frac{1}{(-i \omega + 1/\tau_{\rm imp})} \label{ds1}
\end{equation}
where the system is characterized by the thermodynamic state variables $\varepsilon$, the energy density, and $P$, the pressure--we will specify their temperature
dependencies shortly in Eq.~(\ref{csy}). The factor of $(2e)^2$ is a consequence of our choice
for the normalization of $\rho$; note that product $2e\rho$ measures the net charge density, and so is
independent of this convention.
We assume there is a dilute concentration of impurities which relax the conserved
momentum\footnote{Umklapp scattering can also lead to momentum relaxation, but this is exponentially small at low $T$.}, and lead to the scattering rate $1/\tau_{\rm imp}$---the temperature dependence of $\tau_{\rm imp}$ is specified in Eq.~(\ref{taures}). The parameter $v$ is a velocity characteristic of the
quantum critical point which is assumed to have dynamic critical exponent
$z=1$. Finally, the crucial parameter $\sigma_Q$, is the same quantum conductance which appeared in Eq.~(\ref{ds}).
However, as one moves away from the critical coupling
$g=g_c$ and adds a non-zero $\mu$ in Fig.~\ref{phasediag2},
$\Phi_\sigma$ will acquire a dependence upon the ratios $(g-g_c)/T^{1/\nu}$ and $\mu/T$ which can be included
unchanged in our
results below (here $\nu$ is the usual correlation length exponent).

It is perhaps helpful to note here the ``non-relativistic'' limit
of Eq.~(\ref{ds1}), which does not constitute the regime of primary
interest of this paper. In this limit $\varepsilon + P \approx |\rho| m v^2$ (where
$m$ is the mass of the particles), and then the second term takes the
form of the conventional Drude result.

To develop a theory for the Nernst effect, we need to apply a magnetic field $B$ to the system described so far. A central result of this paper is that for not too strong $B$ fields, the
Nernst response, and a set of related thermoelectric
transport co-efficients, are completely determined by the thermodynamic variables and impurity scattering rate appearing in Eq.~(\ref{ds1}) and the single universal transport co-efficient $\sigma_Q$. In particular, {\em no additional\/} transport co-efficients are needed. Thus there are a large number of Wiedemann-Franz-like relations which relate all the thermoelectric response functions to the regular part of the electrical
conductivity in zero field, $\sigma_Q$.
We will also determine the
frequency dependence of these transport co-efficients;
explicit expressions are given below.

In their work,  Bhaseen {\em et al.\/} \cite{bgs} only considered a non-zero $B$, with $\rho=1/\tau_{\rm imp}=0$. Their primary new result concerned the longitudinal
thermal
conductivity, $\kappa_{xx}$ at zero frequency. Our result for $\kappa_{xx}(\omega=0)$ is consistent with
theirs, and further, we show further that it is related to $\sigma_Q$ by a Wiedemann-Franz like identity
(Eq.~(\ref{kappaxxres2})
below). However, remarkably, unlike the conventional identity which specifies the ratio of
$\kappa_{xx}$ to the electrical {\em conductivity\/},  our identity specifies the ratio
of $\kappa_{xx}$ to the electrical {\em resistivity.} This suggests a physical picture of transport
currents carried by vortices rather than particles, at least when the perturbation associated with $B$
is larger than that associated with $\rho$.

\subsection{Characterization of systems under consideration}
\label{sec:systems}

Let us now specify the class of theories to which our results
apply. Current theories of the superfluid-insulator transition
in non-random systems in 2+1 dimensions are described by quantum
field theories which are Lorentz invariant, and are therefore
conformal field theories (CFTs). Consequently, we will mainly
restrict our attention here to $T>0$ ``quantum critical'' phases
of CFTs, and the general structure of their response to a non-zero $\rho$
(which is not restricted to be small) and small $B$ and $1/\tau_{\rm imp}$. We expect that many of our
results, and especially the magnetohydrodynamic analysis in
Section~\ref{sec:mhd}, can be generalized to a wider class of
systems, but we will not discuss such a generalization here.

Specific examples of CFTs to which our results apply are:\\
({\em i\/}) The superfluid-insulator transition of the boson
Hubbard model on a two-dimensional lattice with a density of an
integer number of bosons per unit cell. The bosons carry charge
$\pm 2e$ because they are caricatures of Cooper pairs. The critical
point is described \cite{fwgf} by the Wilson-Fisher fixed point of
the $|
\psi|^4$ field theory of a complex scalar
$\psi$ (representing the boson annihilation operator), see Eq.~(\ref{spsi}) below.
This field theory also has
a dual representation \cite{peskin,dasgupta}
in terms of a vortex field $\varphi$ coupled to an emergent U(1) gauge
field. Our results apply equally to both representations, and the observable properties do not
depend, naturally, on whether the particle or vortex representation is used to describe
the CFT. \\
({\em ii\/})  The superfluid-insulator transition of the boson
Hubbard model on a two-dimensional lattice with a mean density of
a rational number, $p/q$ (with $p$, $q$, co-prime integers), of
bosons per unit cell.
A `deconfined' critical point \cite{senthil}
is then possible \cite {balents1,balents2} between the superfluid
and an insulator with valence-bond-solid order and is described by
the theory of $q$ flavors of vortex fields $\varphi_{\ell}$
coupled to an emergent U(1) gauge field. This field theory can
also be `undualized' to a ``quiver gauge theory'' of
fractionalized
bosons with charge $ \pm 2e/q$ \cite{balents1}.\\
({\em iii\/}) Electronic models with a $d$-wave superconducting
ground state can also undergo deconfined phase transitions to
insulating states with valence-bond-solid order \cite
{courtney,balents3}. The CFTs of these transitions have Dirac
fermion degrees of freedom, representing the gapless, Bogoliubov
quasiparticle excitations of the $d$-wave superconductor, in
addition to the multiple vortex and gauge fields found above in ({\em ii\/}).\\
({\em iv\/}) Yang-Mills gauge theories with a SU($N$) gauge group
and $\mathcal{N}=8$ supersymmetry. These are attracted in the
infrared to a superconformal field theory (SCFT) which is solvable
in the large $N$ limit via the AdS/CFT mapping. This solution has
been used in previous work \cite{herzog,m2cft} to obtain the
collisionless-to-hydrodynamic crossover in the transport of a
conserved SO(8) R-charge, as well as an exact value for $\Phi_\sigma$.
Here we will examine, as in other recent work \cite{hartnoll}, the
deformation of the SCFT by a non-zero $B$ and $\rho$. The $B$
field and density $\rho$ are both associated with a U(1) subgroup
of the SO(8) R charge. After the AdS/CFT mapping, $B$ and
$\rho$ correspond to the magnetic and the electric charge of a black hole
in AdS space. We will present exact results for the
conserved current correlators of the dyonic black hole in
Section~\ref{sec:dyon}, which allows us to obtain corresponding exact
results for the Nernst and related thermoelectric responses of the
SCFT. In the appropriate hydrodynamic limit, these results are
found to be in full agreement with the more general
magnetohydrodynamic analysis in Section~\ref{sec:mhd}. Additional
comparisons between the hydrodynamic and AdS/CFT results
appear in a separate paper \cite{hh}.

It is worth reiterating that not all of the above CFTs are purely bosonic,
and the examples in ({\em iii\/}) and ({\em iv\/}) contain fermionic
degrees of freedom. Furthermore, in cases ({\em ii\/}) and ({\em iii\/}), the
bosonic degrees of freedom of the CFT are not Cooper pairs, but fractions of a Cooper pair
with charges determined by the density of the Mott insulator.

\subsection{A simple model}
Before presenting our main results, it is useful to establish notation by explicitly writing down the
simplest of the CFTs listed above. This is the $|\psi|^4$ field theory for bosons with charge $\pm 2e$
and action
\begin{equation}
\mathcal{S} = \int d^2 r d \tau \left[ \left| \left(\partial_\tau - i \frac{2e}{\hbar} A_\tau \right)
\psi \right|^2 + v^2 \left| \left(\vec{\nabla} - i \frac{2e}{\hbar c} \vec{A} \right)  \psi \right|^2
- g |\psi|^2 + \frac{u}{2} |\psi|^4 \right] \,, \label{spsi}
\end{equation}
where $\vec{r} = (x,y)$ is a 2-dimensional spatial co-ordinate, $\tau$ is imaginary time, $g$ is the
coupling which tunes the system from the superfluid to the insulator (see Figs.~\ref{phasediag},\ref{phasediag2}),
and the quartic coupling $u$ is attracted to the Wilson-Fisher fixed point value in the infrared limit
associated with the CFT. The electromagnetic gauge potential $A_\mu$ is {\em non-fluctuating}
(and is not to be confused with
the emergent U(1) gauge field of the vortex CFTs noted above).
Its time-component takes an imaginary value (in imaginary time) which determines the chemical
potential
\begin{equation}
i 2e A_\tau = \mu \,,
\end{equation}
while the spatial components take $\tau$-independent values so that
\begin{equation}
\vec{\nabla} \times \vec{A} = B,
\end{equation}
with a spacetime-independent magnetic field $B$. The density, $\rho$, is defined, as usual, by the
derivative of the partition function with respect to the chemical potential
\begin{equation}
\rho =   \frac{k_B T}{\hbar \mathcal{V}} \left\langle \frac{\partial \mathcal{S}}{\partial \mu}\right\rangle\, ,
\end{equation}
where $\mathcal{V}$ is the volume of the system. We reiterate that $\rho$ measures the {\em
difference\/} in the density from that of the
commensurate, $T=0$, insulating state, and not the total density. Also, $\rho$ is a charge density in the sense that it measures the number density of particles minus
the number density of anti-particles.

Another parameter above which will be important for experimental comparisons is the velocity $v$.
Note that it plays the role of the velocity of ``light'' in the ``relativistic'' CFT. It is determined
here by the parameters of the underlying boson Hubbard (or other microscopic) model
whose superfluid-insulator transition is described by the above CFT. It is important to distinguish $v$ from the velocity $c$, which is the actual velocity of light. Here $c$ merely plays the role of a
coupling constant which relates the value of $B$ to physical CGS units, and is not a velocity
associated with the dynamics of the physical model under consideration. Because $v \ll c$, we can
neglect the actual relativistic corrections associated with the physical quantum fluctuations of the photon field $A_\mu$.

With the definition of $v$ at hand, we can now begin comparing the
various energy scales which characterize the system. The largest
energy scales which characterize the deviation from the $T=0$
quantum critical point are $k_B T$, an energy scale $m_0 \sim
|g-g_c|^\nu$ associated with the deviation from critical coupling,
and the chemical potential $\mu$.
We will generally assume that $k_B T$ is the largest of these
scales; our results allow $m_0$ and $\mu$ to be of order $k_B T$, but
not too much larger. For the solvable SCFT theories considered in
Section~\ref{sec:dyon}, the energy scales associated with
$\rho$ and $B$ will not be restricted to small values. However,
for the more general analysis in the remainder of the paper which
applies also to non-supersymmetric CFTs, we will assume that the
perturbation due to $B$ is small, and in particular,
\begin{equation}
\hbar v \sqrt{2eB/(\hbar c)} \ll k_B T.
\label{small}
\end{equation}

Some thermodynamic state variables will also appear in our transport result.
Their temperature dependencies obey scaling forms
similar to those computed earlier for the present theory at $\rho=0$ in Ref.~\onlinecite{csy},
and for $\rho \neq 0$ in Ref.~\onlinecite{zphys}.
In particular, we will need results for the energy density, $\varepsilon$,
and the pressure, $P$ which obey
\begin{eqnarray}
\varepsilon &=& k_B T \left( \frac{k_B T}{\hbar v} \right)^2 \Phi_\varepsilon, \nonumber \\
P &=& k_B T \left( \frac{k_B T}{\hbar v} \right)^2 \Phi_P
\label{csy}
\,,
\end{eqnarray}
where, as in Eq.~(\ref{ds}), $\Phi_{\varepsilon,P}$ are dimensionless universal
numbers which depend on the ratios $(g-g_c) /(k_B
T)^{1/\nu}$ and $\mu/(k_B T)$ \cite{csy,zphys}. The dependence on $B$ is not difficult to account
for, but will be subdominant, and non-singular, under the
condition in Eq.~(\ref{small}).

The final parameter to introduce in our theory of the Nernst effect and other thermoelectric response functions is the momentum relaxation rate $1/\tau_{\rm imp}$. The theory defined so far conserves
total momentum, and thus, such relaxation requires an additional perturbation. We assume that the
relaxation is caused by a weak random impurity potential $V(r)$ which couples to the most relevant
perturbation allowed by symmetry at the CFT fixed point.
For the present theory, this is the operator $|\psi|^2$, and therefore the impurity action is
\begin{equation}
\mathcal{S}_{\rm imp} = \int d \tau \int d^2 r V(r) |\psi (r, \tau) |^2 .
\label{simp}
\end{equation}
We will take a ``quenched'' average over the ensemble of impurity potentials which obey
\begin{equation}
\overline{V(r)} = 0~~;~~\overline{ V(r) V(r') } = V_{\rm imp}^2 \delta^2 (r-r'),
\label{vaverage}
\end{equation}
and work to order $V_{\rm imp}^2$. Note that total energy and charge are conserved in the
presence of $V(r)$, and momentum is the only conserved quantity which will relax. We estimate the
momentum
relaxation rate in Section~\ref{sec:mom} and find
\begin{equation}
\frac{1}{\tau_{\rm imp}} \sim V_{\rm imp}^2 T^{3-2/\nu}. \label{taures}
\end{equation}
The condition for this scattering to be
small is
\begin{equation}
\hbar/\tau_{\rm imp} \ll k_B T.
\label{tau}
\end{equation}
The present model has $\nu \approx 2/3$, and so $1/\tau_{\rm imp}$ depends upon temperature
only very weakly. Indeed, all the CFTs noted earlier are expected to have a similar value for $\nu$. It is therefore a reasonable first approximation to treat $1/\tau_{\rm imp}$ as a temperature-independent
constant.
We will also ignore the dependence of $1/\tau_{\rm imp}$ on $B$ and $\rho$,
under the condition in Eq.~(\ref{small}).

\subsection{Summary of results for the thermo-electric response}
We finally turn to a statement of our main results for the transport co-efficients. We are interested
in the
response of the electrical current $\vec{J}$ and the heat current $\vec{Q}$ to an applied electric
field $\vec{E}$ and a temperature gradient $\vec{\nabla} T$. The precise definitions of $\vec{J}$ and $
\vec{Q}$
appear in the contexts of the models studied in the body of the paper. The electric field can be
applied
by allowing for a weak spatial dependence in the chemical potential $\mu$ (which is then, formally,
the electrochemical potential) with $2e \vec{E} = -\vec{\nabla} \mu$, while the temperature
gradient describes a similar weak spatial dependence in $T$. The transport co-efficients are defined by the relation
\begin{equation}
\left( \begin{array}{c} \vec{J} \\ \vec{Q} \end{array} \right) =
\left( \begin{array}{cc} \hat{\sigma} & \hat{\alpha} \\  T \hat{\alpha} & \hat{\overline{\kappa}} \end{array} \right)
\left( \begin{array}{c} \vec{E} \\ -\vec{\nabla} T \end{array} \right),
\label{alltrans}
\end{equation}
where $\hat{\sigma}$, $\hat{\alpha}$ and $\hat{\overline{\kappa}}$ are $2 \times 2$
matrices acting on the spatial indices $x,y$.
Rotational invariance in the plane imposes the form
\bea
\hat{\sigma}=\sigma_{xx}\, \hat{1} +\sigma_{xy}\hat{\epsilon},
\eea
where $\hat{1}$ is the identity, and $\hat{\epsilon}$ is the antisymmetric tensor $\hat{\epsilon}_{xy}=-\hat{\epsilon}_{yx}=1$. $\sigma_{xx}$ and $\sigma_{xy}$ describe the longitudinal and Hall conductivity, respectively.
An analogous form holds for the thermoelectric
conductivity $\hat{\alpha}$ (which determines the Peltier, Seebeck, and Nernst
effects), as well as for the matrix $\hat{\overline{\kappa}}$ which governs thermal transport in the absence of electric fields. The latter applies to samples connected to conducting leads, allowing for a stationary current flow.
In contrast, the thermal conductivity, $\hat{\kappa}$, is defined as the heat current
response to $-\vec{\nabla} T$ in the absence of an electric
current (electrically isolated boundaries). It is given by
\begin{equation}
\hat{\kappa} = \hat{\overline{\kappa}} -T \hat{\alpha} \hat{\sigma}^{-1} \hat{\alpha}. \label{kappadef}
\end{equation}
Finally, the Nernst response is defined as the electric field induced by a
thermal gradient in the absence of an electric current, and is given in linear response by the relation $\vec{E}=-\hat{\vartheta}\vec{\nabla}T$, with
\begin{equation}
\hat{\vartheta} = - \hat{\sigma}^{-1} \hat{\alpha}. \label{nernstdef}
\end{equation}
The Nernst signal is the transverse response, $e_N \equiv \vartheta_{yx}$. The Nernst {\em co-efficient} is usually defined as $\nu=e_N/B$, which tends to become field independent at small $B$. The Nernst signal is expected to be positive if it is due to driven vortices, while it is generally negative if it arises from quasiparticle excitations~\cite{oganesyan}.

We now present our main results for the transport co-efficients. For the computations using AdS/CFT
applied to the SYM theory in Section~\ref{sec:dyon}, results can be obtained for general
external frequency, $\omega$. However, our more general hydrodynamic results apply only for
$\hbar \omega \ll k_BT$, and this condition is assumed in the remainder
of this section. We begin by presenting our complete result for the frequency
dependence of the longitudinal electrical conductivity (whose $B\rightarrow 0$
limit was already quoted in Eq.~(\ref{ds1})):
\begin{equation}
\sigma_{xx} = \sigma_Q \left[ \frac{(\omega+ i/\tau_{\rm imp})
( \omega + i \gamma + i \omega_c^2 /\gamma + i /\tau_{\rm imp})}{(\omega + i\gamma + i/
\tau_{\rm imp})^2 - \omega_c^2}
\right] \,.
\label{sigmaxxres}
\end{equation}
The overall scale is set by the quantum conductance $\sigma_Q$ introduced in Eq.~(\ref{ds1}), and
the remainder depends upon two important frequency
scales which will appear throughout our analysis. These frequencies are
\begin{equation}
\omega_c \equiv \frac{2 e B \rho v^2}{c (\varepsilon+P)},
\label{omegacres}
\end{equation}
and
\begin{equation}
\gamma \equiv \frac{\sigma_Q B^2 v^2}{c^2 (\varepsilon+P)}. \label{gammares}
\end{equation}
We identify the first frequency, $\omega_c$, as a cyclotron frequency. This seems a natural
interpretation in view of the damped resonance present in the denominator of Eq.~(\ref{sigmaxxres}). Note that  in the non-
relativistic limit  where $ \varepsilon + P \approx |\rho| m v^2$, $\omega_c$ reduces to the familiar result $\omega_c=2eB/(mc)$. For relativistic particles the cyclotron frequency decreases with the energy $E$ as $\omega_c(E)=2eBc/E$. In the present context where $v$ plays the role of the velocity of light, this is modified to $\omega_c(E)=2eB/c\cdot v^2/E$.
The hydrodynamic expression~(\ref{omegacres}) can be regarded as a thermal average over cyclotron frequencies $\omega_c(E\sim T)$, while the proportionality to the charge density, $\omega_c\sim \rho$, reflects the fact that particles and antiparticles circle in opposite senses.

We can consider the cyclotron mode as arising either from the motion of particles
and anti-particles, or from the motion of vortices and anti-vortices. In the latter interpretation, the
roles of $B$ and $\rho$ are interchanged, whilst the expression
for $\omega_c$ remains invariant. We will have more to say about this
`dual' interpretation in the body of the paper, and further results appear in a separate paper \cite{hh}.

The second frequency, $\gamma$, is the damping frequency of the cyclotron mode of
particles and anti-particles. Note that this damping is present
even in the absence of external impurities, and is a consequence of collisions
between particles and anti-particles which are executing cyclotron
orbits in opposite directions. This should be contrasted from the behavior
of a Galilean-invariant system ({\em i.e.\/}, a system with no anti-particle
excitations) for which Kohn's theorem \cite{kohn} guarantees an infinitely
sharp cyclotron mode. The sharpness of the cyclotron resonance is determined by the ratio
\bea
\frac{\gamma}{\omega_c}=\Phi_\sigma \frac{B}{\phi_0\rho},
\eea
which up to the factor $\Phi_\sigma$ equals the number of flux quanta,
\begin{equation}
\phi_0 = \frac{hc}{2e},
\end{equation}
applied per charge $2e$.

We will see later that a {\em different\/} frequency plays the role of the damping
of the cyclotron mode when it is interpreted as due to the motion of vortices and anti-vortices. In that case
the damping frequency is
\begin{equation}
\gamma_{v}=\frac{\omega_c^2}{\gamma} = \frac{4e^2 \rho^2 v^2}{\sigma_Q (\varepsilon+P)}
\,.
\label{vortexdamping}
\end{equation}
There is an obvious `dual' structure apparent upon comparing Eq.~(\ref{gammares}) and
(\ref{vortexdamping}), which we will discuss in more
detail.
Note that the cyclotron resonance will be visible only in ultrapure samples where $1/\tau_{\rm imp}\ll \omega_c$. In this case, the cyclotron resonance is sharp in the thermoelectric response associated with particle transport if $\gamma/\omega_c= B/(\phi_0\rho) \ll 1$, while the same condition implies a washed out resonance in the dual response functions associated with vortices. In the opposite regime, $\gamma/\omega_c\gg 1$, the vortex response, and in particular the Nernst effect, should exhibit a sharp cyclotron resonance in ultrapure samples.

Another notable feature of Eq.~(\ref{sigmaxxres}) is the singular nature of the limit associated with
the small perturbations of the quantum critical region
of Fig.~\ref{phasediag2}. In particular, note that for the d.c.
conductivity at $\omega=0$, the value of $\sigma_{xx}$ depends upon
the order of limits of $\rho \rightarrow 0$, $B \rightarrow 0$, and
$1/\tau_{\rm imp} \rightarrow 0$. This singular limit reflects the fact that the low frequency transport studied previously at $\rho=B=0$ in Refs.~\onlinecite
{damle,m2cft} has ballistic energy propagation and an infinite thermal conductivity. For non-zero $
\rho$ or $B$, the energy and number currents can mix with each other, leading to a finite thermal
conductivity and an order unity correction to $\sigma_{xx}$, as anticipated for the case $\rho=0$ in Ref.~\onlinecite{bgs}.

A useful property of Eq.~(\ref{sigmaxxres}) is that it depends only upon the
combination $\omega + i /\tau_{\rm imp}$. This is actually a property of the
long distance limit of the hydrodynamic equations presented in Section~\ref{sec:mhd}, and is
obeyed by all the transport co-efficients. The remainder
of this section will therefore present results only in the d.c. limit, while
the $\omega$ dependence can be easily reconstructed by replacing $1/\tau_{\rm imp} \rightarrow 1/\tau_
{\rm imp} - i \omega$ (as long as $\hbar \omega \ll
k_B T$).

\subsection{Nernst effect}
\label{sec:Nernst}

Our central result for the Nernst signal 
is
\begin{eqnarray}
e_N=\vartheta_{yx} &=& \left( \frac{k_B}{2e} \right) \left(
\frac{\varepsilon+P}{k_B T\rho} \right) \left[ \frac{\omega_c /\tau_{\rm imp}}{(\omega_c^2/\gamma + 1/
\tau_{\rm imp})^2 + \omega_c^2}\right]
\label{nernstres1} \\
&=&  \frac{1}{\Phi_\sigma} \left( \frac{k_B}{2e} \right) \left(
\frac{\varepsilon+P}{k_B T B/\phi_0} \right) \left[ \frac{\gamma/\tau_{\rm imp}}{(\omega_c^2/\gamma + 1/
\tau_{\rm imp})^2 + \omega_c^2}\right] ,
\label{nernstres2}
\end{eqnarray}
where $\Phi_{\sigma}$ is the dimensionless universal number
appearing in the expression for the conductivity $\sigma_Q$ in Eq.~(\ref{ds}).
We have expressed the Nernst signal in terms of its quantized unit,
\begin{equation}
\frac{k_B}{2e} = 43.086~\mbox{$\mu$V/K},
\end{equation}
times dimensionless ratios in the various brackets.
We can use the relation $\varepsilon + P \approx T s$, where
$s$ is the entropy density (see Eq.~(\ref{thermo})), valid for small $\mu$, $\rho$, to identify the
factor in the second brackets as approximately the entropy per particle in Eq.~(\ref{nernstres1}),
and as the entropy per vortex in Eq.~(\ref{nernstres2}).
The combination of Eqs.~(\ref{ds}), (\ref{csy}), (\ref{taures}) and (\ref{nernstres1},\ref
{nernstres2}) now implies an interesting and non-trivial dependence of the Nernst signal
on $B$ and $T$. Those should be observable in experimental regimes where the entire thermoelectric response is dominated by critical superconducting fluctuations, as will be discussed further in Section~\ref{sec:expts}.

\subsection{Other thermo-electric transport coefficients}
We conclude this introductory section by mentioning two other results for transport co-efficients
whose limiting forms can be compared with earlier computations. For the transverse thermo-electric
conductivity we obtain
\begin{eqnarray}
\alpha_{xy}
&=&  \left( \frac{2ek_B}{h} \right) \left(
\frac{s/k_B}{B/\phi_0} \right) \left[\frac{
\gamma^2 + \omega_c^2 + \gamma/\tau_{\rm imp}(1 - \mu\rho/(Ts))}{(\gamma + 1/\tau_{\rm imp})^2 +
\omega_c^2}\right] \,. \label{alphaxyres}
\end{eqnarray}
While in most recent experiments, the electric conductivity $\hat\sigma$ receives the largest contribution from non-critical carriers, the thermoelectric conductivity is dominated by superconducting fluctuations, even far above $T_c$. It is thus the main quantity to be compared with recent experimental observations in Section~\ref{sec:expts}.
As with earlier results, $\alpha_{xy}$ has been written in terms of the quantum unit of the
thermoelectric co-efficient \cite{podolsky,behnia1}, $2 e k_B/h=6.7$~nA/K, times dimensionless ratios.
In the absence of impurity scattering, $1/\tau_{\rm imp} \rightarrow 0$,
the factor in the square brackets  is unity, and we have
$\alpha_{xy} = sc/B$, a result obtained long ago for non-interacting fermions~\cite{obra,sillin,streda} and later derived by Cooper
{\em et al.} \cite{cooper} for interacting fermions, and by Bhaseen {\em et al.} \cite{bgs}
for the superfluid-insulator transition.

For the longitudinal thermal conductivity we obtain
\begin{eqnarray}
\kappa_{xx} &=&  \Phi_{\sigma} \left( \frac{ k_B^2 T}{h} \right) \left(
\frac{\varepsilon+P}{k_B T \rho} \right)^2 \left[ \frac{(\omega_c^2/\gamma)(\omega_c^2/\gamma+ 1/\tau_
{\rm imp})}{(\omega_c^2/\gamma + 1/\tau_{\rm imp})^2 + \omega_c^2}\right] \label
{kappaxxres1} \\
&=& \frac{1}{\Phi_\sigma} \left( \frac{k_B^2 T}{h} \right) \left(
\frac{\varepsilon+P}{k_B T B/\phi_0} \right)^2 \left[ \frac{\gamma(\omega_c^2/\gamma+ 1/\tau_{\rm imp})}
{(\omega_c^2/\gamma + 1/\tau_{\rm imp})^2 + \omega_c^2}\right],
\label{kappaxxres2}
\end{eqnarray}
where now $k_B^2 T/h$ is the quantum unit of thermal conductance.
In the limit $1/\tau_{\rm imp} \rightarrow 0$ and $B \rightarrow 0$, the factor  within the square
brackets in Eq.~(\ref{kappaxxres1}) reduces to unity. The resulting  expression for $\kappa_{xx}$
relates it to $\sigma_Q$ in a Wiedemann-Franz-like relation, as has been noted by Landau and
Lifshitz  \cite{ll} (and elaborated on recently in Ref.~\onlinecite{sonstarinets}). This relation
suggests
a physical picture of transport due to particles/anti-particles carrying
charges $\pm 2 e$ and entropy per particle $s/\rho$.

In the complementary limit of  $1/\tau_{\rm imp} \rightarrow 0$ and $\rho \rightarrow 0$
the factor within the square brackets in Eq.~(\ref{kappaxxres2}) reduces to unity. Now $\kappa_
{xx}$ is proportional to the resistivity $1/\sigma_Q$, indicating a picture of transport due to
vortices of net density $B/\phi_0$.
The value of $\kappa_{xx}$
has the same dependence upon all parameters as that obtained by Bhaseen {\em et al.} \cite{bgs}. We can also compare the value of the numerical prefactor. For $\Phi_\sigma$ we use the value $1.037/\epsilon^2$ obtained in
Ref.~\onlinecite{damle} in the $\epsilon$-expansion ($\epsilon=3-d$ where $d$ is the spatial dimension), which is the same expansion by Bhaseen {\em et al.} \cite{bgs}. It is also easy to compute the value of $\Phi_\varepsilon+\Phi_P$ in the same expansion: to the leading order needed, these are just given by
the values for free, massless, relativistic bosons in $d=3$, which yields
$\Phi_\varepsilon + \Phi_P = 4 \pi^2 /45 + \mathcal{O} (3-d)$.
Using these values we obtain
the same result for $\kappa_{xx}$ as in Eq.~(24) of Ref.~\onlinecite{bgs}, with their dimensionless
parameter $g=4.66$ (not to be confused with our coupling $g$). This is to be compared with their value $g=5.55$.
The origin of this numerical discrepancy is not clear to us.
We believe Eq.~(\ref{kappaxxres2}) is an exact identity in $d=2$, but it is possible that it is modified
when $d$ is close to 3.

The outline of the paper is as follows. Section~\ref{sec:expts} will compare the result for the Nernst effect and the thermoelectric response Eq.~(\ref{alphaxyres}) with experiments on the cuprate superconductors
and on Nb$_{0.15}$Si$_{0.85}$ films. Section~\ref{sec:mhd} will present the derivation of these results using a hydrodynamic analysis of transport near a generic, 2+1 dimensional, ``relativistic'' quantum critical
point perturbed by a chemical potential, a magnetic field, and weak impurity scattering. An estimation of the impurity
scattering rate appears in Section~\ref{sec:mom}. Section~\ref{sec:dyon} will describe the exact solution for
transport near a supersymmetric quantum critical point, perturbed by a chemical potential and a magnetic field,
which is solvable by the AdS/CFT mapping to the physics of a dyonic black hole in 3+1 spacetime dimensions.
Some technical details appear in the appendices.

\section{Comparison with experiments}
\label{sec:expts}

Our main results for the Nernst signal have already been stated in Section~\ref{sec:Nernst}.
In the following subsections, we will these results with recent observations in the cuprate superconductors \cite{ong1,ong2,ong3,ong4},
and also briefly discuss experiments in Nb$_{0.15}$Si$_{0.85}$ films \cite{behnia1,behnia2}.
As mentioned before, in most of these experiments the electrical conductivity is dominated by non-critical fermionic contributions which are not captured by our relativistic hydrodynamic description.
On the other hand, the transverse thermo-electric response $\alpha_{xy}$ is expected to be predominantly due to superconducting fluctuations and the vortex liquid. In practice $\alpha_{xy}$ is conveniently measure via the Nernst signal, using the relation $\alpha_{xy}\approx \sigma_{xx}\vartheta_{yx}$. The latter holds if the non-critical Hall conductivity is small, $\sigma_{xy}\ll \sigma_{xx}$, as is usually the case.

It is convenient to perform the experimental comparisons by rescaling $B$ and $\rho$ so that they are both measured in units of (energy)$^2$,
\begin{equation}
B = \overline{B} \phi_0/(\hbar v)^2~~~;~~~\rho = \overline{\rho} /(\hbar v)^2.
\end{equation}
Further we observe that in typical experiments the flux per (excess) particle is very small, $B/\rho\ll 1$ and therefore $\gamma/\omega_c\ll 1$.
In this regime Eq.~(\ref{alphaxyres}) simplifies to
\bea
\left( \frac{h}{2e k_B} \right) \alpha_{xy} &\approx& \frac{s/k_B}{B/\phi_0}(\tau_{\rm imp}\omega_c)^2 \frac{1+\gamma/(\tau_{\rm imp}\omega_c^2)(1-\mu\rho/sT)} {1+(\tau_{\rm imp}\omega_c)^2} \label{alphaplot1}\\
&\approx &\Phi_s \overline{B}\,(k_BT)^2\left(\frac{2\pi\tau_{\rm imp}}{\hbar}\right)^2\frac{\overline{\rho}^2+\Phi_\sigma\Phi_{\varepsilon+P}(k_BT)^3\,\hbar/2\pi\tau_{\rm imp}}{\Phi_{\varepsilon+P}^2(k_B T)^6+\overline{B}^2\overline{\rho}^2(2\pi\tau_{\rm imp}/\hbar)^2},
\label{alphaplot2}
\eea
where in the second line we have assumed a fully relativistic regime with $s\sim T^2$ and $\varepsilon+P\sim T^3$, and $\mu\rho\ll sT$.
We recall that $\Phi_{\varepsilon+P}$ and $\Phi_\sigma$ are universal functions of $\mu/T$, and have an additional dependence on $(g-g_c)/T^{1/\nu}$.

\subsection{The cuprates}

Given the relative simplicity of our model of the cuprate superconductors, detailed quantitative comparisons with the observations of Ref.~\onlinecite{ong4} are probably premature. In particular, we have omitted the
influence of long-range Coulomb interactions, which modifies the spectrum of boson density fluctuations,
and likely leads to a superfluid-insulator quantum critical point which is not Lorentz invariant \cite{fg,jinwu}. Also, although Dirac fermion excitations are included in some of the CFTs mentioned above
(corresponding to the nodal points of the $d$-wave superconductor) other Fermi excitations
associated with a Fermi surface may also be important, especially in the case of NbSi.
Keeping these caveats in mind, it is nevertheless useful to examine the extent to which
the present model can describe the observations. As we will now show, using physically reasonable values of the parameters in the theory, our results describe the overall
absolute magnitude of the observations, and numerous qualitative trends \cite{ong4} remarkably well.

We work here with a simple caricature of our predictions: We ignore the $T$ and $\rho$ dependence of the universal functions and simply
treat them as constants, $\Phi_{\sigma}\approx 1.037$, $\Phi_{\varepsilon+P}^{(2d)}\approx \Phi_{s}^{(2d)}\approx 3\zeta(3)/\pi\approx 1.148$. This is equivalent to assuming in Figs.~\ref{phasediag},\ref{phasediag2} that $g=g_c$ and $\mu=\rho=0$ for the purpose of evaluating these functions. It is not difficult to extend our theory to include the influence of these
corrections to the leading quantum-critical behavior, but such a detailed analysis would not be commensurate
with the other simplifications noted above.

We notice that for small $B$, Eqs.~(\ref{alphaplot1},\ref{alphaplot2}) predict a Nernst signal linear in $B$. At not too large temperatures, the second term in the numerator of (\ref{alphaplot2}) can be neglected and the ratio $\alpha_{xy}/B$ is seen to increase with decreasing temperature as $1/T^4$,
\bea
\label{T4}
\frac{\alpha_{xy}}{B}(B\rightarrow 0)
&\approx &\left( \frac{2e k_B}{h\phi_0} \right)\frac{\Phi_s}{\Phi_{\varepsilon+P}^2} \left(\frac{2\pi\tau_{\rm imp}}{\hbar}\right)^2\frac{\rho^2 (\hbar v)^6}{(k_B T)^4}.
\eea
Such a power law with exponent $4$ was indeed observed  over two orders of magnitude in signal strength in
underdoped La$_{2-\delta}$Sr$_\delta$ CuO$_4$ (LSCO, $\delta\leq 0.12$), cf., Fig.~4 in Ref.~\onlinecite{podolsky}.
Assuming a typical doping $\delta-\delta_I=-0.025$ for underdoped LSCO with a lattice constant $a=3.78$ \AA, we obtain a constraint for $\tau_{\rm imp} v^3$ from fitting (\ref{T4}) to the experimental value $\alpha_{xy}/B=0.48/(T/30{\rm K})^4$~nA/KT per layer~\cite{podolsky}. Assuming a typical scattering time $\tau_{\rm imp}\approx 10^{-12}$s, 
we obtain
an estimate for the velocity $\hbar v\approx 47$ meV\AA. These are reasonable parameter values, with the velocity $v$ being of the order of the characteristic velocity found in Ref.~\onlinecite{bala}.



\begin{figure}[t]
\centering \includegraphics[width=4in]{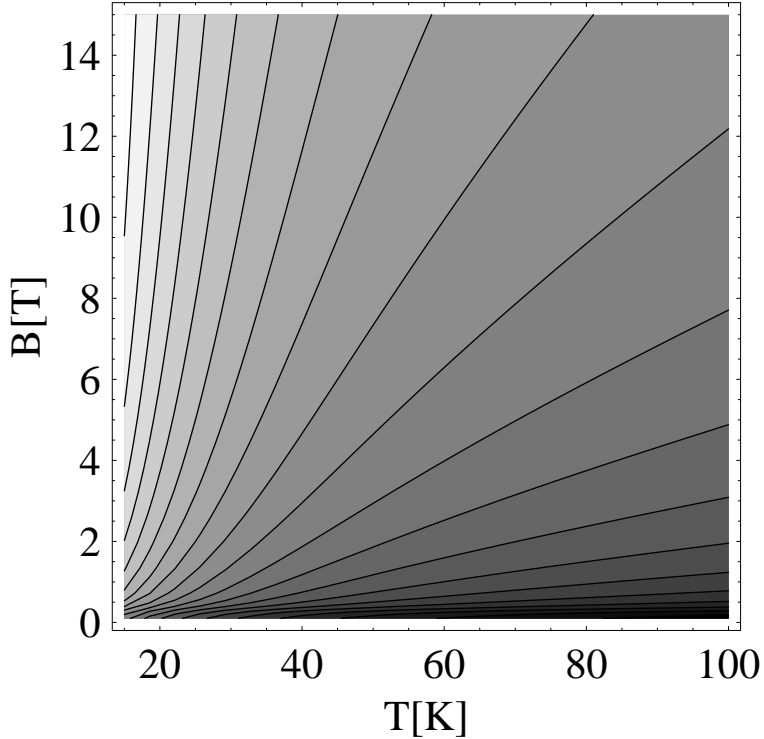}
\caption{Contour plot (with logarithmic spacing) of the thermoelectric conductivity $\alpha_{xy}$ (Eq.~\ref{alphaplot2}) as a function of temperature $T$ and magnetic field $B$, for parameters $\hbar v=47$ meV \AA, $\delta-\delta_I=0.025$ and $\tau_{\rm imp}=10^{-12}$s estimated for LSCO. In the ordered low temperature regime $T<T_c\approx 30$K, Eq.~(\ref{alphaplot2}) will receive modifications.
}
\label{alphacontour}
\end{figure}

The result of Eq.~(\ref{alphaplot1}) is plotted as function of both $T$ and $B$ in Fig.~\ref{alphacontour}.
This contour plot should be compared, {\em e.g.}, with Fig. 13 in Ref.~\onlinecite{ong4} in the underdoped regime, above the superconducting transition $T_c$.

Using the above parameter estimates we predict the cyclotron resonance
\bea
\omega_c= 6.2 {\rm GHz}\cdot \frac{B}{1{\rm T}}\left(\frac{35{\rm K}}{T}\right)^3,
\eea
which, at $T=35$K, is by a factor $\omega_c/\omega_c^{({\rm el})}=2 m^{({\rm el})} v^2\cdot \rho/(\varepsilon+P)\approx 0.035$ smaller than the cyclotron frequency of free electrons. However, as mentioned before, this resonance can only be observed in ultrapure samples where $1/\tau_{\rm imp}\ll \omega_c$, which is clearly not the case in LSCO.

\begin{figure}[t]
\centering \includegraphics[width=4in]{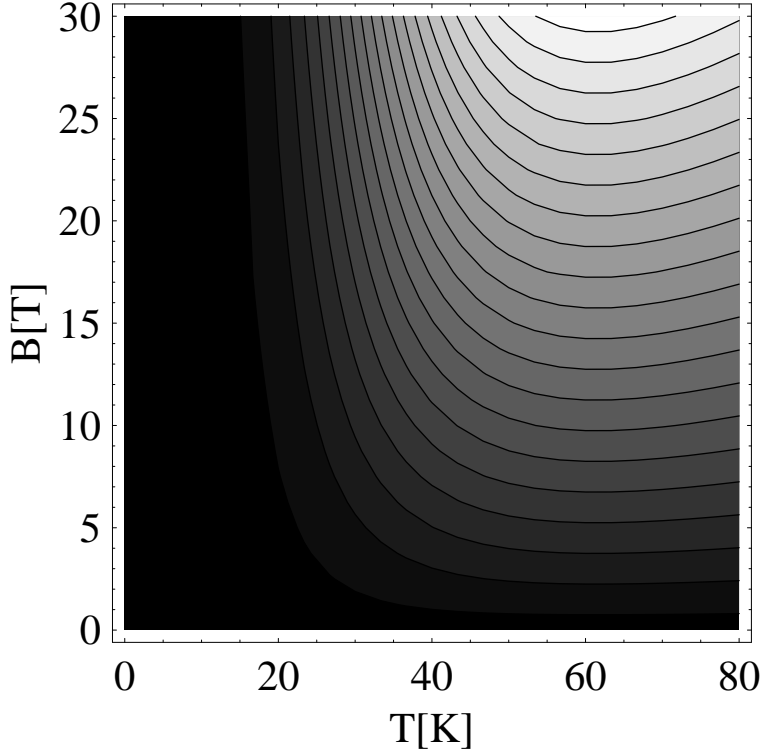}
\caption{Contour plot (linear scale) of the Nernst signal $e_N=\vartheta_{yx}$ (Eq.~\ref{nernstplot}) close to a quantum critical point, as a function of temperature $T$ and magnetic field $B$. The parameters are the same as for Fig.~\ref{alphacontour}. The signal strength in the plot ranges up to $10\mu$V/K.
}
\label{Nernstcontour}
\end{figure}

Having estimated the velocity $v$ and the scattering rate $\tau_{\rm imp}$, we can make a quantitative prediction for the Nernst signal in the vicinity of a quantum critical point where the entire thermo-electric response is expected to be dominated by critical fluctuations. In this case, Eq.~(\ref{nernstres1}) can be cast into the form
\begin{equation}
 \vartheta_{yx} = \left( \frac{k_B}{2e} \right)\frac{ (\Phi_\varepsilon+\Phi_P)^2 \Phi_\sigma^2 \overline{B} (k_B T)^5 (\hbar/(2 \pi \tau_{\rm imp}))}
{ \Bigl[ \overline{\rho}^2 + \Phi_\sigma (\Phi_\varepsilon + \Phi_P) (k_B T)^3 (\hbar/(2 \pi \tau_{\rm imp})) \Bigr]^2
+ \Phi_\sigma^2 \overline{\rho}^2 \overline{B}^2} \,, \label{nernstplot}
\end{equation}
which is plotted in Fig.~\ref{Nernstcontour}

\subsubsection{Hall resistance}
\label{sec:hall}

Very recently, measurements of the Hall resistance in the high field normal state of 
YBa$_2$Cu$_3$O$_{6.5}$ have been reported \cite{taillefer}. The focus of the authors was on magnetoresistance
oscillations; these oscillations are quantum interference effects which cannot be reproduced by the effective classical
hydrodynamic models employed here (under the condition in Eq.~(\ref{small})), and so are beyond the scope
of the present paper. However, the authors also reported a background Hall resistance which, surprisingly, was {\em negative\/}.
The sample has hole density $\delta=0.1$. As argued in Section~\ref{sec:intro}, the density of mobile carriers, $\rho$, which appears in the hydrodynamic theory \cite{balents1} (and which contributes a Magnus force on vortices \cite{lorenz}) is given by the {\em difference} in density between
the superconductor and the proximate Mott insulator. Using an insulator at
$\delta_I = 0.125$,  we have $\rho = -0.025/(2 a^2)$. This
negative value of $\rho$ 
provides a very natural explanation of the observed negative Hall resistance. Also, we can predict that the Hall resistance should change sign as $\delta$ is increased beyond $\delta_I$.

We can make a more quantitative comparison with experiments. In Eq.~(\ref{fres2}) in Section~\ref{sec:linear}, we report the value of the Hall resistivity, $\rho_{xy}$, and the zero frequency limit of that result is
\begin{equation}
\rho_{xy} = \frac{B}{2e\rho c} \left[ 1 - \frac{(1/\tau_{\rm imp})^2}{(1/\tau_{\rm imp} + \omega_c^2/\gamma)^2 + \omega_c^2} \right]\,;
\label{rhoxy}
\end{equation}
in the absence of impurity scattering ($\tau_{\rm imp} \rightarrow \infty$),
this result was noted in Ref.~\onlinecite{hartnoll}.
Using the value of $\rho$ noted above, at $B= 60$T, we determine that the prefactor of the square brackets
in Eq.~(\ref{rhoxy}) is -2.2 k$\Omega$. For the factor within the square brackets, we assume the same parameters as found above for LSCO, and conclude that it is close to unity.
This result is to be compared with the observed resistance per layer \cite{taillefer}
at this field of -3.9 k$\Omega$, which is quite reasonable agreement for this simple model. 

\subsection{NbSi}
We also note the experiments on amorphous films of Nb$_{0.15}$Si$_{0.85}$ reported in Refs.~\onlinecite{behnia1,behnia2}
The normal phase, $T>T_c$, of these films exhibits a number of features that are
consistent with our hydrodynamic results, when taken to their non-relativistic limit.
In particular, $\alpha_{xy}$ as given in Eq.~(\ref{alphaplot1})
displays a functional dependence on magnetic field akin to that reported in Ref.~\onlinecite{behnia2}:
\begin{equation}
\label{Behniaobs}
\alpha_{xy}\propto \frac{B}{\xi^{-4}+\ell_B^{-4}}\propto \frac{B}{1+(B/B_0)^2},
\end{equation}
with $B_0=\hbar c/e\xi^2$, which was interpreted as the physics being controlled by the shorter of the superconducting coherence length $\xi$ and the magnetic length $\ell_B=(\hbar c/eB)^{1/2}$~\cite{behnia2}.
We mention that the low $B$ data, i.e., the Nernst coefficient $\nu=\lim_{B\rightarrow 0}\alpha_{xy}/B\sigma_{xx}$ measured in Ref.~\onlinecite{behnia1}, was very successfully described by the theory of Gaussian fluctuations~\cite{ussishkin}. However, the crossover (\ref{Behniaobs}) and the high field behavior $\alpha_{xy}\sim 1/B$ remained unexplained in earlier theories. Our magnetohydrodynamic approach may give a hint to the origin of the latter.
We believe that the similarity of Eqs.~(\ref{alphaplot1}) and (\ref{Behniaobs}) is not a mere coincidence. Rather, it leads us to speculate that the scattering time $\tau_{\rm imp}$ should be identified with
\bea
\tau_{\rm imp}= \frac{B}{B_0\omega_c}=\frac{m \xi^2}{\hbar} \approx  \left(k_F\ell\right)\, \frac{\xi^2}{v_F\ell} \sim \tau_{\rm GL},
\eea
the Ginzburg-Landau life time of fluctuating Cooper pairs~\cite{larkin05}.
Here we have used the free electron value (non-relativistic limit) for the cyclotron frequency $\omega_c^{({\rm el})}=eB/m^{({\rm el})}c$. Further, $v_F = \hbar k_F/m$ is the Fermi velocity, and we have used that $k_F \ell\approx {\cal O}(1)$ in the studied amorphous NbSi~\cite{behnia1}. The estimate $\tau_{\rm imp}\approx \tau_{\rm GL}$  suggests that the suppression of the Nernst signal at high fields is due to the Cooper pairs starting to perform entire orbits over their life time.

We may use the above guess of $\tau_{\rm imp}$ to express the low field limit of $\alpha_{xy}$, as
\bea
\alpha_{xy}(B\ll B_0)=\frac{k_B e}{\hbar}\frac{\xi^2}{\ell_B^2}\frac{s\xi^2}{k_B},
\eea
where we have invoked a small value of $\gamma$ to approximate the last numerator in Eq.~(\ref{alphaplot1}) by to $1$.
It is interesting to note that apart from the last factor which describes the entropy per coherence volume, this expression has the same parameter dependence as the one derived from Gaussian fluctuations in Ref.~\onlinecite{ussishkin}.


\section{Magnetohydrodynamics}
\label{sec:mhd}

The remainder of this paper will revert to natural dimensionless units
with $\hbar=k_B =2e=v=1$, and absorb a factor of $1/c$ in the definition of $B$.

Here we will focus on the nature of quantum critical transport in the hydrodynamic region \cite
{damle,m2cft} where $\hbar \omega \ll k_B T$. The condition in Eq.~(\ref{small}) ensures $\hbar \omega_c \ll k_B T$, and a relativistic formulation is appropriate if also $m_0 \lesssim k_B T$ is satisfied. We will use the method described by Landau and
Lifshitz~\cite{ll}, which was recently reviewed in the context of a string theory computation~\cite{sonstarinets}. These previous analyses were carried out for $B=0$ and $1/\tau_{\rm imp} = 0$, and only considered the longitudinal electrical and
thermal conductivities.
Here we will show how the hydrodynamic analysis can be extended to include non-zero values
of these parameters. Further, we will obtain results for the frequency dependence of the full set of transport
co-efficients in Eq.~(\ref{alltrans}). These results are
consistent with the exact results obtained via AdS/CFT for a particular SCFT which are presented in Section~\ref{sec:dyon}
and Ref.~\onlinecite{hh}---the latter results however extend over a wider regime of parameters.

The fundamental ingredients of a hydrodynamic analysis are the conserved
quantities and their equations of motion. Unlike in the theory of dynamics near classical, finite temperature critical points \cite{halphoh}, here we do not need to explicitly consider the order parameter dynamics for the effective equations of motion of the low frequency theory. The key difference is that $k_B T/\hbar$ constitutes an intrinsic relaxation time for the order parameter fluctuations, and we are only interested in much lower frequency scales. In contrast, at a classical critical point, the relaxation time diverges.
The frequency scales larger than $k_B T/\hbar$ cannot be addressed by the
methods below, and require a full quantum treatment of the dynamics of the CFT.

The conserved quantities of interest
 are the electrical charge, the energy, and the momentum. For the relativistic theories under consideration, these can be assembled into the electrical current 3-vector~\footnote{Notice that upon restoring $v$, the 3-current reads $J^\mu = (\rho v,0,0)$.} $J^\mu=(\rho,J^x,J^y)$, and the
stress-energy tensor $T^{\mu\nu}$. We will use standard relativistic notation
with the metric tensor $g^{\mu\nu} = \mbox{diag}(-1,1,1)$ and coordinates $x^\mu=(v t,x,y)$. For the moment, we will ignore the momentum relaxation due to the weak impurity potential
in Eq.~(\ref{simp}), and include its effects shortly below. With total momentum conserved, the equations of motion obeyed by the total electrical, momentum and energy currents are
\begin{eqnarray}
\partial_\mu J^{({\rm tot})\,\mu} &=& 0, \label{e1p}  \\
\partial_\nu T^{({\rm tot})\,\mu\nu} &=& F^{\mu \nu}J^{({\rm tot})}_\nu .  \label{e1}
\end{eqnarray}
The first equation represents the conservation of charge and requires no further comment. The second equation describes the evolution of the stress energy tensor, and the term on the right hand side represent the effects of the external $B$ field.
Here $F^{\mu \nu}$ is the applied magnetic field which takes the fixed value~\footnote{More precisely, if we also allow for an electrical field in the lab frame and restore $v$, the field tensor takes the form
\begin{equation}
F^{\mu \nu} = \left( \begin{array}{ccc} 0 & (c/v)E_x & (c/v)E_y \\
-(c/v)E_x & 0 & B \\
-(c/v)E_y & -B & 0
\end{array} \right).
\end{equation}
The equation of motion (\ref{e1}) is a priori valid for velocities $v'\ll v$, where the non-relativistic limit applies. However, it is valid as a relativistically covariant equation under Lorentz transformations $\Lambda^\mu_\nu$ characterized by the limit velocity $v$ ({\em not} $c$), if it is understood that $F^{\mu \nu}$ transforms as an antisymmetric tensor, $F^{\mu \nu}=\Lambda^\mu_\alpha\Lambda^\nu_\beta F^{\alpha\beta}$ under a change of reference frame. This is exact up to negligible corrections of order $O([v/c]^2)$.
}
\begin{equation}
F^{\mu \nu} = \left( \begin{array}{ccc} 0 & 0 & 0 \\
0 & 0 & B \\
0 & -B & 0
\end{array} \right), \label{fmn}
\end{equation}
and the right hand side of Eq.~(\ref{e1}) describes the Lorentz force exerted by this field, as discussed, {\em e.g.\/}, in Ref.~\onlinecite{llem}.
In equilibrium, we have $J^{({\rm tot})\,\mu} = (\rho,0,0)$ and then the term proportional
to $F^{\mu \nu}$ vanishes, as expected.

To use the equations (\ref{e1p}) and (\ref{e1}), we need to relate $J^{({\rm tot})\,\mu}$ and $T^{({\rm tot})\,\mu\nu}$ to parameters which define the local thermodynamic equilibrium, and a three-velocity
$u^\mu$ which represents the
velocity of the system in local equilibrium with respect to the laboratory frame.
As usual \cite{ll}, the three-velocity $u^\mu \equiv dx^\mu/d\tau$ satisfies $u^\mu
u_\mu = -1$, and $u^\mu = (1,0,0)$ in the equilibrium frame where there is no energy flow~\footnote{Note that the absence of an energy flow in the rest frame {\em defines} the velocity vector $u^\mu$ in the "dynamic" frame used throughout this paper. Alternatively, one can formulate the hydrodynamics in the kinetic frame~\cite{demian} where $\overline{u}^\mu$ denotes the velocity associated with the electrical current via $J^\mu=\rho \overline{u}^\mu$. However, the dynamic frame is the natural frame to work with, since we assume local equilibrium. $u^i$ is canonically conjugate to the momentum density $T^{0i}$, and the energy current has a natural expression in terms of $u^\mu$ (cf.~Eq.~(\ref{heat})).
}.
For the thermodynamic parameters we will use the charge density, $\rho$, the energy density $\varepsilon$, the pressure $P$, and the magnetization density $M$; we define the pressure, $P$, as the negative of the grand potential per unit volume,
and $M$ as the derivative of the latter with respect to $B$.

Using these parameters the stress energy tensor of a fluid is given by
\begin{eqnarray}
T^{(tot)\,\mu\nu} &=& T^{\mu\nu} - M^{\mu\gamma}F^\nu_{\ \gamma} +T^{E\mu\nu},\nn\\
T^{\mu\nu} &=& (\varepsilon+P) u^\mu u^\nu + P g^{\mu\nu} +
\tau^{\mu\nu},
\label{e0}
\end{eqnarray}
where
\begin{equation}
M^{\mu \nu} = \left( \begin{array}{ccc} 0 & 0 & 0 \\
0 & 0 & M \\
0 & -M & 0
\end{array} \right), 
\end{equation}
is the polarization tensor~\cite{degroot}. (The electric polarizations $M^{0i}=-M^{i0}$ vanish in the absence of electric fields in the lab frame.)

The electrical current is given by
\begin{eqnarray}
J^{({\rm tot})\,\mu} &=& J^\mu + \partial_\nu M^{\mu \nu}, \label{Jmu} \\
J^\mu &=& \rho u^\mu  + \nu^\mu. \label{e00}
\end{eqnarray}

The 'dissipative current' $\nu^\mu$ accounts for the fact that the charge current and the energy current are not simply proportional to each other. This is because there is a heat flow even in the absence of matter convection, which is a consequence of particle-antiparticle creation and annihilation.

We have introduced the {\em transport currents} \cite{cooper}, $J^\mu$ and   $T^{\mu\nu}$ which represent observable quantities which can couple to probes external to the system. The remaining contribution
to $J^{({\rm tot})\mu}$ is the magnetization current \cite{ussishkin,podolsky,cooper,streda}, which is induced due to spatial variations in the local magnetization density.
The coupling of the magnetization to the magnetic field contributes an extra contribution $- M^{\mu\gamma}F^\nu_{\ \gamma}$ to the stress energy tensor, reducing its spatial diagonal to $P_{\rm int}=P-MB$ (see also Appendix~\ref{app:mag}).
The origin and the physics of this term has also
been discussed by Cooper {\em et al} \cite{cooper}. Finally $T^{E\mu\nu}$ represents the ``energy magnetization current''. We will not need an explicit
expression for this quantity here, apart from the fact that it obeys
$\partial_\nu T^{E \mu\nu} =0$. Expressions will be given later in the paper when
we consider specific CFTs: for the super Yang Mills theory in Section~\ref{sec:dyon}, and for the Wilson-Fisher
fixed point in Appendix~\ref{app:mag}.
With these magnetization currents subtracted out, the residual transport currents continue to obey the equations of motion as in Eqs.~(\ref{e1p}) and (\ref{e1}):
\begin{eqnarray}
\partial_\mu J^\mu &=& 0, \nonumber \\
\partial_\nu T^{\mu\nu} &=& F^{\mu \nu} J_\nu.  \label{e1t}
\end{eqnarray}

In the expressions for the transport currents in Eqs.~(\ref{e0}) and (\ref{e00}), we assume that $\varepsilon$, $P$, $\rho$, and
$M$ are thermodynamic functions of the local chemical potential,
$\mu$, the temperature $T$, and the magnetic field $B$. We
will treat $u^\mu$, $\mu$, and $T$ as the ``independent'' degrees of freedom
which respond to external perturbations, and assume that the remaining
thermodynamic variables will follow according to the equation of state.
In equilibrium,
the non-zero components of $J^\mu$ and $T^{\mu\nu}$ are
\begin{equation}
J^t = \rho~~;~~T^{tt} = \varepsilon~~;~~T^{xx}=T^{yy}=P-MB.
\label{tpint}
\end{equation}
Eqs.~(\ref{e0}) and (\ref{e00}) also contain
the dissipative components of the stress-energy tensor and the current, as introduced in Ref.~\onlinecite{ll}; these are
are orthogonal to $u^\mu$
\begin{equation}
u_\mu \tau^{\mu\nu} = u_\mu \nu^\mu = 0, \label{e2}
\end{equation}
and will be determined below by imposing the requirement that the total entropy increases under time evolution.

We are now in a position to introduce the scattering due to a dilute concentration of impurities. We assume that their dominant effect is to introduce a relaxation in the local transport momentum density: impurity scattering conserves charge and energy, and we do not expect the magnetization currents to relax by impurity scattering. Thus, we modify Eqs.~(\ref{e1t}) to
\begin{eqnarray}
\partial_\mu J^\mu &=& 0, \nonumber \\
\partial_\nu T^{\mu\nu} &=& F^{\mu \nu} J_\nu
+ \frac{1}{\tau_{\rm imp}} \left( \delta^{\mu}_\nu + u^\mu u_\nu \right)
T^{\nu\gamma} u_\gamma .  \label{e1tt}
\end{eqnarray}
The new term in the second equation in Eq.~(\ref{e1tt})
represents the impurity scattering. The impurities, as described by the random potential in Eq.~(\ref{simp}) are assumed to be at rest in the laboratory frame. The projection operators built out of the $u^\mu$ in the second term in Eq.~(\ref{e1}) ensure that in the laboratory frame only the total momentum, {\em i.e.\/}, $T^{i0}$, is relaxed. We will discuss a computation of the value of $\tau_{\rm imp}$ later in Section~\ref{sec:mom}.

Following Landau and Lifshitz \cite{ll}, we now use the positivity
of the entropy production to constrain the expression for the dissipative
components $\nu^\mu$ and $\tau^{\mu\nu}$.  First we notice from
Eq.~(\ref{e0}) that
\begin{equation}\label{udT}
   u_\nu \d_\mu T^{\mu\nu} = - (\varepsilon+P)\d_\mu u^\mu
   - u^\mu \d_\mu \varepsilon + u_\nu \d_\mu \tau^{\mu\nu} ,
\end{equation}
and from Eq.~(\ref{e1tt}) that
\begin{equation}
u_\nu \d_\mu T^{\mu\nu} = F^{\mu\nu} u_\mu \nu_\nu.
\label{e3}
\end{equation}
Using the thermodynamic relations
\begin{equation}
   \varepsilon+P = Ts + \mu \rho, \qquad d\varepsilon = Tds + \mu d\rho, \label{thermo}
\end{equation}
Eq.~(\ref{e2}) and current conservation, Eqs.~(\ref{udT}) and
(\ref{e3}) can be transformed into
\begin{equation}
   \d_\mu (su^\mu) = \frac\mu T \d_\mu \nu^\mu - \frac{1}{T} F^{\mu\nu} u_\mu \nu_\nu
    - \frac{\tau^{\mu\nu}}T \d_\mu u_\nu,
\end{equation}
or
\begin{equation}
\label{entropy}
   \d_\mu \left(su^\mu - \frac\mu T\nu^\mu\right) =
   \nu^\mu \left[\frac{1}{T}\left(-\d_\mu \mu + F^{\mu\nu} u_\nu\right) +\mu \frac{\d_\mu T}{T^2}\right] - \frac{\tau^{\mu\nu}}
T \d_\mu u_\nu\,.
\end{equation}
It is natural to interpret the left hand side as the divergence of the entropy current. Accordingly we can interpret the three vector
\bea
\label{heat}
Q^\mu = sT\,u_d^\mu - \mu \nu^\mu=(\varepsilon +P) u^\mu-\mu J^\mu \equiv J^{E\mu}-\mu J^\mu
\eea
as the heat current. We have also introduced the energy current $J^{E\mu}=(\varepsilon +P) u^\mu$.

Since the entropy can only increase, the right hand side of (\ref{entropy}) must be positive.  Generalizing the arguments of Landau
and Lifshitz, we deduce the most general expressions for the dissipative currents which are linear in spatial gradients and the velocity,
\begin{align}
   \nu^\mu &=  \sigma_Q (g^{\mu\nu}+u^\mu u^\nu)\left[\left(-\d_\nu \mu + F_{\nu\lambda}u^{\lambda}\right) +\mu \frac{\d_\mu T}{T}\right],\label{nu-constit}\\
   \tau^{\mu\nu} &= -(g^{\mu\lambda}+ u^\mu u^\lambda)\left[ \eta (\d_\lambda u^\nu + \d^\nu u_{\lambda})
   +\left(\zeta-\eta\right)
    \delta^\nu_\lambda \d_\alpha u^\alpha\right]\,.
\end{align}
Here $\eta$ and $\zeta$ are the shear and bulk viscosities, and
$\sigma_Q$ is a conductivity.
Notice that there are only three independent
transport co-efficients. We will neglect velocity gradients for the most part
in this paper, and so the viscosities do not appear in our main results.
Consequently, we have the remarkable feature that all transport response functions
depend only upon a single dissipative transport co-efficient $\sigma_Q$.
Notice that in the dissipative current the gradient of the chemical potential appears in combination with the electromagnetic forces $F^{\mu\nu}u_\nu$, which is natural since it is equivalent to an electric field.

\subsection{Linear response}
\label{sec:linear}

We will now follow the strategy of Kadanoff and Martin \cite{km}:
Use the equations of hydrodynamics
to solve the initial value problem in linear response, and compare
the results to those obtained from the Kubo formula in order to extract transport coefficients and their frequency dependence.

First, we address the solution of the initial value problem in
hydrodynamics. We begin by choosing our independent variables: from
the structure of the above equations, it appears convenient to
choose the four variables $T$, $\mu$, and $u^x$ and $u^y$. So we write
\begin{eqnarray}
\mu (r, t) &=& \mu + \delta \mu (r,t),
\nonumber \\
T (r, t) &=& T + \delta T (r,t) , \label{pert}
\end{eqnarray}
while $F^{\mu\nu}$ is fixed at the value in Eq.~(\ref{fmn}). We also
write $u^\mu$ as
\begin{equation}
u^\mu = \left( \begin{array}{ccc} 1 \\
v_x (r,t) \\
v_y (r,t)
\end{array} \right) , \label{u}
\end{equation}
where $v_x$, $v_y$ are of the same order as $\delta\mu$ and $\delta
T$.

The other variables,  $\varepsilon$, $P$, and $\rho$ are constrained by
local thermodynamic equilibrium to have the form
\begin{eqnarray}
\rho (r,t) = \rho + \delta \rho &\equiv & \rho +  \left.
\frac{\partial \rho}{\partial \mu} \right|_{T} \delta \mu + \left.
\frac{\partial \rho}{\partial T} \right|_{\mu} \delta T,
\nonumber \\
\varepsilon (r,t) = \varepsilon + \delta \varepsilon &\equiv& \varepsilon +
\left. \frac{\partial \varepsilon}{\partial \mu} \right|_{T} \delta \mu
+ \left. \frac{\partial \varepsilon}{\partial T} \right|_{\mu} \delta T,
\nonumber \\
P (r,t) = P + \delta P &\equiv& P +  \rho \delta \mu + s \delta T. \label{thermoder}
\end{eqnarray}
The various components of the
stress-energy tensor and the current vector are perturbed accordingly. To
linear order we have
\begin{eqnarray}
\delta T^{tt} &=& \delta \varepsilon, \nonumber \\
\delta T^{ti} &=& T^{ti}=(\varepsilon + P)v_i, \nonumber \\
\delta T^{ij} &=& \delta P \delta_{ij}  -  \eta (\partial_i v_j +
\partial_j v_i - \delta_{ij}
\partial_k v_k) -\zeta \delta_{ij} \partial_k v_k, \nonumber \\
\delta J^t &=& \delta \rho, \nonumber \\
\delta{\vec{J}} &=& {\vec{J}}=\rho \vec{v}+\vec{\nu},\nonumber\\
\vec{\nu} &=&  \sigma_Q \left(  -\vec{\nabla}\mu +\vec{v}\times \vec{B} +\mu\frac{\vec{\nabla}T}{T} \right).
\label{perturbs1}
\end{eqnarray}

For small perturbations the conservation laws take the form
\bea
\label{chargecons}
\partial_t \rho+\vec{\nabla}{\vec{J}}&=&0,\\
\label{econs}
\partial_t \varepsilon+\vec{\nabla}\vec{J}^E&=&0,\\
\partial_t \vec{J}^E&=&-\vec{\nabla}P-\eta \vec{\nabla}^2 \vec{v}-\zeta \vec{\nabla}(\vec{\nabla}\cdot \vec{v})+{\vec{J}}\times\vec{B},
\label{momentumcons}
\eea
with the energy and heat currents
\bea
\vec{J}^E&=&(\varepsilon+P)\vec{v},\\
\vec{Q}&=&\vec{J}^E-\mu {\vec{J}}=(\varepsilon +P) \vec{v}-\rho\mu\vec{v}-\mu \vec{\nu}=sT\, \vec{v}-\mu \vec{\nu},
\eea
The crucial remnant of the relativistic theory in the linearized hydrodynamics is the fact that the energy and particle currents are in general {\em not} parallel,
\bea
\label{currents}
\vec{J}^E&=& \frac{\varepsilon+P}{\rho}\left[{\vec{J}}-\sigma_Q\left(  -\vec{\nabla}\mu +\vec{v}\times \vec{B} +\mu\frac{\vec{\nabla}T}{T} \right)\right] \nn\\
&=& \frac{\varepsilon+P}{\rho}{\vec{J}} +\frac{(\varepsilon+P)^2}{T\rho^2}\sigma_Q
\vec{\nabla}T+\frac{\varepsilon+P}{\rho^2}\sigma_Q\left(
-\vec{\nabla}P+\vec{J}\times \vec{B}\right),
\eea
where we have used (\ref{thermo}) to rewrite the dissipative current. The energy current consists of three parts: the first two are familiar from non-relativistic theory as the convection of matter and heat flow due to a thermal gradient, with a thermal conductivity~\cite{ll} $\overline{\kappa}=  \sigma_Q(\varepsilon+P)^2/(T\rho^2)$. The last term in (\ref{currents}) is proportional to the acceleration vector, and is a purely relativistic phenomenon~\cite{Gravitation}. One can easily see that this term is responsible for the damping $\gamma$ of the cyclotron mode (cf.,~Eq.~(\ref{gammares})), by using the above relation to substitute for ${\vec{J}}$ in the momentum conservation law (\ref{momentumcons}).

To complete the hydrodynamic analysis, we solve the equations
(\ref{chargecons})-(\ref{momentumcons}) for arbitrary initial
values $\delta T^0$, $\delta \mu^0$ and $v_x^0=v_y^0=0$, and
obtain the response in the electrical current $\vec{J}$ and the
heat current $\vec{Q}$. The 'heat density' associated to the
latter, $q(r)=\varepsilon(r)-\mu \rho(r)$ is canonically conjugate
to the temperature at fixed chemical potential~\cite{km}.

After a Fourier transform in space and a Laplace transform in time,
the linear response of any quantity $A$ obeys
\bea
A(\vec{k},\omega) &=& \frac{G_{A ; \varepsilon-\mu\rho}({\vec k},\omega) -
G_{A; \varepsilon-\mu\rho} ({\vec k},0)}{i \omega} \frac{\delta T^0({\vec k})}{T}+
\frac{G_{A ; \rho}({\vec k},\omega) -
G_{A; \rho} ({\vec k},0)}{i \omega} \delta \mu^0 ({\vec k})\nonumber
\\
&& \quad
 + \frac{G_{A ; T^{0i}}({\vec k},\omega) -
G_{A; T^{0i}} ({\vec k},0)}{i \omega}
\sum_{i={x,y}}  v^i ({\vec k}),
\eea
where the coefficients are related to retarded equilibrium correlation functions, as can be shown from analyzing an adiabatic perturbation~\cite{km} of the form
\bea
\label{pert2}
\delta {\cal H}(t)=-\int dr \left[\delta \mu(r,t) n(r,t) -\frac{\delta T(r,t)}{T}(\varepsilon(r,t)-\mu n(r,t))-\sum_i v^i(r,t) T^{0i}(r,t)\right].
\eea

Finally, using the conservation laws in the form
\bea
i\omega \rho(\vec{k})&=&i\vec{k}\vec{J}(\vec{k}),\nn\\
i\omega (\varepsilon(\vec{k})-\mu \rho(\vec{k}))&=&i\vec{k}\left[ \vec{J}^E(\vec{k})-\mu\vec{J}(\vec{k})\right]=i\vec{k}\vec{Q}(\vec{k}),
\eea
we obtain
\bea
\label{Kubo}
A(\vec{k},\omega) &=& -\frac{1}{i\omega}\left[\frac{G_{A ; \vec{Q}}({\vec k},\omega) -
G_{A; \vec{Q}} ({\vec k},0)}{i \omega} \left(-\frac{\vec{\nabla}T^0({\vec k})}{T}\right) +\frac{G_{A ; \vec{J}}({\vec k},\omega) -
G_{A; \vec{J}} ({\vec k},0)}{i \omega} \vec{E}({\vec k})\right]\nn
\\
&&  \qquad
 + \frac{G_{A ; T^{0i}}({\vec k},\omega) -
G_{A; T^{0i}} ({\vec k},0)}{i \omega}
\sum_{i={x,y}}  v^i ({\vec k}).
\eea
For $A={\vec J}$ and $A={\vec Q}$ one recognizes the co-efficients of $\vec{E}\equiv -\vec{\nabla}\mu^0$ and $(-\vec{\nabla}T^0/T)$ as $(-1/i\omega)$ times the Kubo formulae for the thermoelectric co-efficients $\hat{\sigma},\hat{\alpha},\hat{\kappa}$. The response to an initial velocity perturbation could be used to extract frequency dependent viscous response functions.

After a Laplace transform in time,
the Eqs.~(\ref{chargecons}) and (\ref{econs}) take the form
\begin{eqnarray}
 i \left(\left. \frac{\partial \rho}{\partial \mu} \right|_{T} \delta \mu^0
+ \left. \frac{\partial \rho}{\partial T} \right|_{\mu} \delta T^0\right) &=& \w \left(\left. \frac{\partial \rho}{\partial \mu} \right|_{T} \delta \mu + \left. \frac{\partial \rho}{\partial T} \right|_{\mu} \delta T\right) +i\vec{\nabla} \left(\rho \vec{v} +
B \sigma_Q \hat{\epsilon}\vec{v} \right)
- i  \sigma_Q T \nabla^2 \left( \frac{\mu}{T}\right) \,,\nonumber \\
 i \left( \left. \frac{\partial \varepsilon}{\partial \mu} \right|_{T} \delta \mu^0 + \left. \frac{\partial \varepsilon}{\partial T} \right|_{\mu} \delta
T^0\right) &=& \w \left(\left. \frac{\partial \varepsilon}{\partial \mu} \right|_{T} \delta \mu + \left. \frac{\partial \varepsilon}{\partial T} \right|_{\mu} \delta T\right) +i (\varepsilon+P)\vec{\nabla}\cdot \vec{v} \,,
\end{eqnarray}
for charge and energy conservation, and Eq.~(\ref{momentumcons})
\begin{eqnarray}
 i(\varepsilon + P )\vec{v}^0  &=& (\w + i/\tau_{\rm imp})(\vep + P) \vec{v} + i(\rho \vec{\nabla}\mu+s\vec{\nabla}T) \nn\\
&~&~~~~~~
-i B \rho \hat{\epsilon} \vec{v}+i B \sigma_Q \hat{\epsilon} (\vec{\nabla}\mu-\frac{\mu}{T} \vec{\nabla}T)+iB^2\sigma_Q\vec{v}\nn\\
&~&~~~~~~ +i\eta \nabla^2 \vec{v}+i\zeta \vec{\nabla}(\vec{\nabla}\cdot\vec{v})\, ,
\label{mhd}
\end{eqnarray}
for momentum conservation.

In the case of weak enough momentum relaxation, the response functions will exhibit peaks associated with the normal modes of these linearized equations. Apart from the damped cyclotron mode discussed above, one finds two diffusive modes, as analyzed in Appendix~\ref{app:modes}. However, those will not be of importance below since we are restricting to long wavelengths in the sequel.

In the limit $k\rightarrow 0$, the current and energy conservation impose $\delta \mu = i\delta\mu^0/\omega$ and $\delta T = i\delta T^0/\omega$, expressing that the decay of initial perturbations can be neglected. Further, the contributions from viscosity can be neglected in this limit.
Upon injection into the momentum conservation equations~(\ref{mhd}), we easily obtain the retarded Greens functions and via the mapping~(\ref{Kubo}) and Kubo formulae the transport co-efficients defined in Eq.~(\ref{alltrans}):
\begin{eqnarray}
\sigma_{xx} 
&=&\sigma_Q \frac{(\omega+i/\tau_{\rm imp})(\omega + i \gamma  +i/\tau_
{\rm imp} + i \omega_c^2/\gamma)}{\left[(\omega + i \gamma + i/\tau_{\rm imp})^2 -
\omega_c^2 \right]},
 \nonumber \\
\sigma_{xy} 
&=&   - \frac{\rho}{B} \frac{( \gamma^2 +  \omega_c^2 - 2 i \gamma \omega + 2
\gamma/\tau_{\rm imp})}{\left[(\omega + i \gamma + i/\tau_{\rm imp})^2 - \omega_c^2
\right]}, \nonumber \\
\alpha_{xx} 
&=&  \frac{s}{B} \frac{\left[ \omega_c (i \omega - 1/\tau_{\rm imp})(1
- (\gamma \mu\rho/(\omega_c^2 Ts))(\gamma + 1/\tau_{\rm imp} - i \omega))\right]}{\left[(\omega + i \gamma
+ i/\tau_{\rm imp})^2 - \omega_c^2 \right]},\nonumber \\
\alpha_{xy} 
&=&   - \frac{s}{B} \frac{\left[\gamma^2 + \omega_c^2  + \gamma (-i \omega + 1/
\tau_{\rm imp})(1-\mu\rho/(Ts))\right]}{\left[(\omega + i \gamma + i/\tau_{\rm imp})^2 - \omega_c^2 \right]},
\nonumber \\
\overline{\kappa}_{xx} 
&=&  -\left(\frac{(\varepsilon+P)^2}{ T B \rho} \right)  \frac{\omega_c \gamma}{\left[(\omega + i
\gamma + i/\tau_{\rm imp})^2 - \omega_c^2 \right]}\nn\\
&~&~~~~~~~~~~~~~ \times
\Biggl\{ 1 +  (1/\tau_{\rm imp} - i \omega) \frac{(s^2 T^2 \omega_c^2 + \gamma^2 \mu^2 \rho^2)}{\gamma \omega_c^2 (\varepsilon+P)^2} + (1/\tau_{\rm imp} - i \omega)^2 \frac{ \mu^2 \rho^2}{ \omega_c^2 (\varepsilon+P)^2} \Biggr\},\nn \\
\overline{\kappa}_{xy} 
&=& -\left(\frac{T s^2}{B \rho}\right) \frac{\omega_c^2}{\left[(\omega + i \gamma + i/\tau_{\rm imp})^2
- \omega_c^2 \right]} \nn\\
&~&~~~~~~~~~~~~~~ \times
 \Biggl\{ 1 - \frac{2\mu \sigma_Q B}{Ts} \frac{( \gamma + 1/\tau_{\rm imp} - i \omega)}{\omega_c} - \left( \frac{\mu \sigma_Q B}{Ts}\right)^2 \Biggr\}. \label{fres1}
\end{eqnarray}

We also computed the thermoelectric co-efficients $\alpha_{xx}$, $\alpha_{xy}$ by examining
the heat current induced by an applied electric field, and
precisely the same result as above was obtained.
This confirms Onsager reciprocity which has to hold since the densities associated with the electric and heat currents are conjugate to $\delta \mu$ and $\delta T/T$, respectively. The validity of Onsager reciprocity is a strong check of the consistency of our hydrodynamic description.

From expressions in Eq.~(\ref{fres1}) we can also derive the
resistivities $\hat{\rho} = \hat{\sigma}^{-1}$, the Nernst
responses defined in Eq.~(\ref{nernstdef}) and the thermal
conductivities defined in Eq.~(\ref{kappadef}).
\begin{eqnarray}
\rho_{xx} &=&  \frac{1}{\sigma_Q} \frac{(\omega+i/\tau_{\rm imp})(\omega + i \gamma  +i/\tau_
{\rm imp} + i \omega_c^2/\gamma)}{\left[(\omega + i \omega_c^2/\gamma + i/\tau_{\rm imp})^2 -
\omega_c^2 \right]}, \nonumber \\
\rho_{xy} &=&   -\frac{B}{\rho} \frac{( (\omega_c^2/\gamma)^2 +  \omega_c^2 - 2 i (\omega_c^2/\gamma)  \omega + 2 (\omega_c^2/\gamma)/\tau_{\rm imp})}{
\left[(\omega + i \omega_c^2/\gamma + i/\tau_{\rm imp})^2 - \omega_c^2 \right]}, \nonumber \\
\vartheta_{xx} &=&  \frac{s}{\rho} \frac{\left[(\omega_c^2/\gamma)^2 +
\omega_c^2  + (\omega_c^2/\gamma) (-i \omega + 1/
\tau_{\rm imp})(1-\mu\rho/(Ts)) - (\mu\rho/(Ts)(-i \omega + 1/\tau_{\rm imp})^2\right]}{\left[(\omega + i \gamma + i/\tau_{\rm imp})^2 - \omega_c^2 \right]},
\nonumber \\
\vartheta_{xy} &=&  - \frac{B}{T} \frac{(i \omega-1/\tau_{\rm imp})}{\left[ (\omega + i \omega_c^2 /
\gamma + i/\tau_{\rm imp})^2 - \omega_c^2\right]}, \nonumber \\
\kappa_{xx} &=&  \frac{(\varepsilon+P)}{T} \frac{(i \omega -1/\tau_{\rm imp}-\omega_c^2/\gamma)}{\left[(\omega + i
\omega_c^2 /\gamma + i/\tau_{\rm imp})^2 - \omega_c^2 \right]}, \nonumber \\
\kappa_{xy} &=& \frac{(\varepsilon+P)}{T} \frac{\omega_c}{\left[(\omega + i \omega_c^2/\gamma+i /\tau_{\rm imp})
^2 - \omega_c^2 \right]}.
\label{fres2}
\end{eqnarray}
These expressions contain the main results that were quoted in Section~\ref{sec:intro}.
Although they appear rather complicated, most of the structure is tightly constrained and the predicted dependencies on $\omega$ are robust consequences of hydrodynamics.

Significant simplifications do appear if in addition to the small $B$ assumption in Eq.~(\ref{small}), we also assume that $\rho$ is small. In particular, let us take
$B \ll T^2$, $\rho \ll T^2$ and $\rho \sim B \sim T^{3/2} \sqrt{\omega}$. Note that in this limit we may simplify Eq.~(\ref{thermo}) to $Ts = \varepsilon + P$. Then, the results in Eqs.~(\ref{fres1}) and (\ref{fres2}) take the following more compact form
(we have set $1/\tau_{\rm imp} = 0$ because, as noted in Section~\ref{sec:intro}, the dependence on impurity scattering is easily restored below by $\omega \rightarrow
\omega + i/\tau_{\rm imp}$):
\begin{eqnarray}
\sigma_{xx} &=&  \sigma_Q \frac{\omega(\omega + i \gamma + i \omega_c^2/\gamma)}{\left[(\omega + i \gamma)^2 - \omega_c^2 \right]}~~~~,~~~~~~~~~~~
\sigma_{xy} =  - \frac{\rho}{B} \frac{( \gamma^2 +  \omega_c^2 - 2 i \gamma \omega)}{\left[(\omega + i \gamma)^2 - \omega_c^2 \right]}, \nonumber \\
\rho_{xx} &=&  \frac{1}{\sigma_Q} \frac{\omega(\omega  + i \omega_c^2/\gamma+ i \gamma)}{\left[(\omega + i \omega_c^2/\gamma)^2 - \omega_c^2 \right]}~~~~,~~~~~~~~~
\rho_{xy}  =    \frac{B}{\rho} \frac{( (\omega_c^2/\gamma)^2 +  \omega_c^2 - 2 i (\omega_c^2/\gamma) \gamma \omega)}{\left[(\omega + i \omega_c^2/\gamma)^2 - \omega_c^2 \right]}, \nonumber \\
\alpha_{xx} &=&  \frac{\rho}{T} \frac{i \omega }{\left[(\omega + i \gamma)^2 - \omega_c^2 \right]} ~~~~,~~~~~~~~~~~~~~~
\alpha_{xy} =   - \frac{s}{B} \frac{\gamma^2 + \omega_c^2 - i \gamma \omega}{\left[(\omega + i \gamma)^2 - \omega_c^2 \right]}, \nonumber \\
\vartheta_{xx} &=&   \frac{s}{\rho} \frac{(\omega_c^2 /\gamma)^2 + \omega_c^2 - i (\omega_c^2/\gamma) \omega}{\left[ (\omega + i \omega_c^2 /\gamma)^2 - \omega_c^2\right]} ~~~~,~~~~
\vartheta_{xy} = -  \frac{B}{T} \frac{i \omega}{\left[ (\omega + i \omega_c^2 /\gamma)^2 - \omega_c^2\right]} ,
\nonumber \\
\overline{\kappa}_{xx} &=&  s \frac{i \omega -\gamma}{\left[(\omega + i \gamma)^2 - \omega_c^2 \right]} ~~~~,~~~~~~~~~~~~~~~~~
\overline{\kappa}_{xy} = -s \frac{\omega_c}{\left[(\omega + i \gamma)^2 - \omega_c^2 \right]}, \nonumber \\
\kappa_{xx} &=&  s \frac{i \omega -\omega_c^2/\gamma}{\left[(\omega + i \omega_c^2 /\gamma)^2 - \omega_c^2 \right]} ~~~~,~~~~~~~~~~~~
\kappa_{xy} = s \frac{\omega_c}{\left[(\omega + i \omega_c^2/\gamma)^2 - \omega_c^2 \right]} .
\label{fres3}
\end{eqnarray}
The above expressions
are now easily observed to obey a remarkable `self-duality' symmetry.
Under the interchanges
\begin{equation}\label{eq:exchange}
\mbox{$\rho \leftrightarrow B$ and $\sigma_Q \leftrightarrow 1/\sigma_Q$},
\end{equation}
the cyclotron frequency $\omega_c$ in Eq.~(\ref{omegacres}) remains invariant, while the damping frequencies in Eqs.~(\ref{gammares},\ref{vortexdamping}) interchange
\begin{equation}
\label{gammas}
\gamma \leftrightarrow \gamma_v=\omega_c^2 /\gamma;
\end{equation}
then note that
the expressions for the transport co-efficients interchange as follows:
\begin{eqnarray}
&&\sigma_{xx},~\sigma_{xy},~ \alpha_{xx},~\alpha_{xy},~\overline{\kappa}_{xx},
~\overline{\kappa}_{xy} \nonumber \\
&&~~~~~~~~~~~~~~~~~\updownarrow \nonumber \\
&&\rho_{xx},-\rho_{xy}, - \vartheta_{xy},- \vartheta_{xx},~\kappa_{xx},-\kappa_{xy}.
\label{dualtrans}
\end{eqnarray}
These relations are consequences of the particle-vortex duality discussed in
Ref.~\onlinecite{m2cft}, and the mapping of the transport co-efficients in Eq.~(\ref{dualtrans}) can be deduced from the mapping $E_i \leftrightarrow \epsilon_{ij} J_j$ in Eq.~(\ref{alltrans}). These duality relations will be discussed further in the context of SCFTs
solvable by AdS/CFT in Section~\ref{sec:dyon} and in Ref.~\onlinecite{hh}:
in this case the duality relations will be found to hold exactly for all $\rho$ and
$B$.

\section{Estimating the momentum relaxation rate}
\label{sec:mom}

As discussed in Section~\ref{sec:intro},
we assume that momentum relaxation is caused by an external perturbation of the
form
\begin{equation}
\mathcal{S}_{\rm imp} = \int d\tau \int d^d x V(x) \mathcal{O} (x,\tau),
\label{v1}
\end{equation}
where $V(x)$ is an external potential which is random function of spatial position $x$, but
independent of $\tau$, with the averages in Eq.~(\ref{vaverage}).
The operator $\mathcal{O}$ is the ``thermal operator'' of the CFT, {\em i.e.\/}, the most relevant
perturbation which drives the CFT massive (despite the name, it has nothing to do with temperature
in the
present context).

We are interested in the modification of the equation of motion of the
momentum density, $T^{i0}$ by the impurity, because this is the only quantity
whose conservation law is spoiled by the presence of impurities. In the absence of other perturbations
from equilibrium we observe from Eq.~(\ref{e1tt}) that the momentum density obeys
\begin{equation}
\frac{\partial T^{it}}{\partial t} = - \frac{1}{\tau_{\rm imp}} T^{it} + \ldots,
\end{equation}
where the ellipses indicates terms which have an explicit spatial gradient and so their spatial
integral vanishes. We will describe here
an estimate of $\tau_{\rm imp}$ to order $V_{\rm imp}^2$.

For definiteness, consider the Wilson-Fisher fixed point of a complex scalar $\psi$ in
Eq.~(\ref{spsi}), although the
argument
easily generalizes to other CFTs. We will also ignore the influence of $B$ and $\mu$ as these are
secondary perturbations. Then $\mathcal{O} = |\psi|^2$ and
\begin{equation}
T^{it} = \varpi^\ast \partial_i \psi + {\rm c.c.}\, ,
\end{equation}
where $\varpi$ is canonical momentum conjugate to $\psi$.
For the following, we need the commutator
\begin{equation}
\Upsilon^j = i [ T^{jt}, \mathcal{O}] = \partial_j (|\psi|^2).
\end{equation}

We now compute $1/\tau_{\rm imp}$ using the memory function method \cite{forster,giam}.
From this approach, the estimate of the momentum relaxation rate is
\begin{equation}
\frac{1}{\tau_{\rm imp}}  = \frac{V_{\rm imp}^2}{\chi_{T}}
\lim_{\omega\rightarrow 0} \frac{1}{\omega} \int \frac{d^d k}{(2 \pi)^d} \mbox{Im} \langle
\Upsilon^i (-k, -\omega) \Upsilon^i (k, \omega) \rangle_{\rm ret}\, ,
\end{equation}
where we are working in general $d$ dimensions, and $\chi_T$ is the momentum density susceptibility {\em i.e.}
\begin{equation}
\chi_T = \int d^d x d\tau \langle T^{it} (x, \tau) T^{it} (0,0) \rangle.
\end{equation}
Using the scaling dimensions
\begin{eqnarray}
\mbox{dim}[T^{i0}] &=& d+1, \nonumber \\
 \mbox{dim}[|\psi|^2] &=& d+1-1/\nu,
\end{eqnarray}
(where $1/\nu$ is the scaling dimension of the coupling conjugate to $\mathcal{O} = |\psi|^2$) we obtain
\begin{eqnarray}
\mbox{dim}[\chi_T] &=& -d-1 + 2~ \mbox{dim}[T^{i0}] = d+1, \nonumber \\
\mbox{dim}\left[ \langle \Upsilon^i (-k, -\omega) \Upsilon^i (k, \omega) \rangle_{\rm ret} \right]
&=&
-d-1 + 2 ~\mbox{dim}[\Upsilon^j] = d+3 -2/\nu .
\end{eqnarray}
Thus $\chi_T \sim T^{d+1}$, and
\begin{equation}
\frac{1}{\tau_{\rm imp}} \sim V_{\rm imp}^2 T^{d+1-2/\nu}, \label{v3}
\end{equation}
which is the result quoted in Eq.~(\ref{taures}) for $d=2$. As noted there, $\nu \approx 2/3$, and so $1/\tau_{\rm imp}$ is
roughly temperature independent.

An alternative, but less constructive argument proceeds along the lines of the discussion
in Ref.~\onlinecite{vbs}.
From Eq.~(\ref{v1}) we have $\mbox{dim}[V] = 1/\nu$, and so from Eq.~(\ref{vaverage})
we have $\mbox{dim}[V_{\rm imp}^2] = -d + 2/\nu$. This is indeed familiar from Harris' criterion which states that disorder is relevant if $\nu < 2/d$. Then, knowing $\mbox{dim} [1/\tau_{\rm imp}]
=1$,
the result (\ref{v3}) follows.

\section{Dyonic black holes}
\label{sec:dyon}

\subsection{AdS$_4$/CFT$_3$ and the black hole solution}

From the point of view of studying quantum critical phenomena, the
AdS/CFT correspondence \cite{Maldacena:1997re} provides a wealth
of new solvable strongly correlated conformal field theories
(CFTs) in 2+1 dimensions. The key feature of these CFTs is that
they admit a large $N$ limit in which they can be described
classically as a gravitational theory in $3+1$ dimensions that
asymptotes to Anti-de Sitter space (AdS$_4$). The CFT is thought
of as living on the `boundary' of the higher dimensional, or
`bulk', spacetime.

The correspondence furthermore allows us to consider departures
from criticality due to finite temperature. This is precisely the
type of systems we are studying in this paper. Finite temperature
in field theory corresponds to allowing the bulk spacetime to
contain a black hole \cite{Witten:1998zw}. The temperature of the
field theory is just the Hawking temperature of the black hole.
Finite temperature dissipation in field theory is dual to bulk
matter fields falling into the black hole.

We wish to consider a CFT with a global $U(1)$ symmetry and a
corresponding charge density $\rho$ and a background magnetic field
$B$. It was explained recently that this is dual to taking a
dyonic black hole, carrying both electric and magnetic charge
\cite{hartnoll}. These black holes are solutions to
Einstein-Maxwell theory in 3+1 dimensions. In this section, and also in a separate
paper \cite{hh}, we will
see how thermoelectric transport properties of the dyonic black hole precisely
agree with our general analysis in the hydrodynamic limit.
The black hole, via the AdS/CFT correspondence, provides a solvable example
of the physics we are studying throughout this paper.
Furthermore, the various particle-vortex dualities we have
discussed above are seen to acquire a very transparent interpretation in
the AdS/CFT correspondence.

The canonical example of the AdS$_4$/CFT$_3$ correspondence
describes the infrared fixed point of maximally supersymmetric
Yang-Mills theory with $SU(N)$ gauge group. The dual gravitational
theory in this case is $M$ theory on AdS$_4 \times S^7$
\cite{Maldacena:1997re}. However, in the large $N$ limit and for
the subset of field theory questions we are asking, this theory
may be consistently truncated to Einstein-Maxwell theory with a
negative cosmological constant in 3+1 dimensions. The dimensional
reduction is performed for instance in \cite{m2cft}.

The action for Einstein-Maxwell theory with a negative
cosmological constant may be taken to be
\be\label{eq:theaction}
I = \frac{\sqrt{2} N^{3/2}}{6 \pi} \int d^4x \sqrt{-g}
\left[-\frac{1}{4} R + \frac{1}{4} F_{\mu\nu} F^{\mu\nu} -
\frac{3}{2} \right] \,,
\ee
which implies the equations of motion
\begin{subequations}
\label{eq:eom}
\bea
R_{\mu\nu} & = & 2 F_{\mu\sigma} F_\nu{}^\sigma -
\frac{1}{2} g_{\mu\nu} F_{\sigma\rho} F^{\sigma\rho} -
3 g_{\mu \nu} \,,
\label{eq:einstein-eqs} \\
\nabla_\mu F^{\mu \nu} & = & 0 \,.
\label{eq:maxwell-eqs}
\eea
\end{subequations}
We have expressed the normalization of the action in terms of the
field theory quantity $N$.

A black hole in $AdS_4$ with planar horizon has metric
\be
ds^2 = \frac{\a^2}{z^2} \left[-f(z) dt^2 + dx^2 + dy^2\right] +
\frac{1}{z^2} \frac{dz^2}{f(z)} \,.
\label{eq:metric}
\ee
The dyonic black hole carries both electric and magnetic charge
\be\label{eq:Ffield}
F = h \a^2 dx \wedge dy + q \a dz \wedge dt \,,
\ee
where $h,q$ and $\a$ are constants that will be related to field
theory quantities shortly. The function appearing in the metric is
then
\be
f(z) = 1 + (h^2 + q^2) z^4 - (1 + h^2 + q^2) z^3 \,.
\label{eq:f}
\ee
In these coordinates, the conformal boundary of the spacetime is
at $z \to 0$ whereas the black hole event horizon is at $z=1$.

In the following subsection we summarize the thermodynamic
properties of this black hole spacetime that were derived in
\cite{hartnoll}, which is also the thermodynamics of the CFT. We
will express thermodynamic quantities in terms of the dual field
theory background magnetic field and chemical potential. These are
related to the constants $q$ and $h$ that appeared in the black
hole solution as follows \cite{hartnoll}
\be\label{eq:charges}
B = h \a^2 \,, \quad \mu = - q \a \,.
\ee

\subsection{The grand canonical ensemble}

We give the thermodynamics of the CFT in terms of the temperature
$T$, the chemical potential $\mu$ and the background magnetic
field $B$. Many variables are most simply expressed in terms of an
auxiliary quantity $\a(T,\mu,B)$ which is determined from
\be
\frac{4 \pi T}{\a} = 3 - \frac{\mu^2}{\a^2} - \frac{B^2}{\a^4} \,.
\ee
The thermodynamic potential is
\be
\Omega = \frac{\sqrt{2} N^{3/2}}{6 \pi} \frac{\mathcal{V} \a^3}{4}
\left(-1 - \frac{\mu^2}{\a^2} + 3 \frac{B^2}{\alpha^4} \right) \,.
\ee
Here $\mathcal{V} = \int\! dx\, dy$ is the spatial volume. The energy
density is
\be
\vep = \frac{\sqrt{2} N^{3/2}}{6 \pi}
\frac{\a^3}{2} \left(1 + \frac{\mu^2}{\a^2} + \frac{B^2}{\a^4} \right) \,.
\ee
The entropy density is
\be
s = \frac{\sqrt{2} N^{3/2}}{6} \a^2 \,.
\ee
The charge density is
\be\label{eq:chargedensity}
\rho =
  \frac{\sqrt{2} N^{3/2}}{6 \pi} \a^2 \frac{\mu}{\a} \,.
\ee
The magnetization is
\be\label{eq:mag}
M = - \frac{1}{\mathcal{V}} \frac{\pa \Omega}{\pa B} = - \frac{\sqrt{2}
N^{3/2}}{6 \pi} \a \frac{B}{\a^2} \,.
\ee
The pressure is
\be
P = M B + \frac{\vep}{2} \,.
\ee
These formulae satisfy the thermodynamic relation
\be
\vep + P = T s + \mu \rho \,.
\ee
Note that the above implies the relation
\be\label{eq:rel1}
\vep = \frac{2}{3} \left(s T + \mu \rho - M B \right) \,.
\ee
Finally, it is useful to define
\be\label{eq:chi}
\chi = \frac{\sqrt{2} N^{3/2}}{6 \pi} T \,.
\ee
Which gives the relation
\be\label{eq:rel2}
\rho M T^2 = - \mu B \chi^2 \,.
\ee

\subsection{Magnetization densities and currents}
\label{sec:dyonmag}

In the following subsection, we will use Kubo formulae to the
obtain transport coefficients of the SCFT from retarded Greens
functions. When applying the Kubo formula to systems with a
background magnetic field, it is necessary to subtract effects due
to magnetization currents, as explained in Cooper {\em et al.\/} \cite{cooper}.
The magnetization currents are
\be
J^{\text{mag.}}_i = \epsilon_{ij} \pa_j M \,,
\ee
and
\be
T^{\text{mag.}}_{ti} = \epsilon_{ij} \pa_j M^E \,.
\ee
Here and below the indices $i,j$ run over the two spatial
coordinates $x$ and $y$. The equilibrium magnetization density $M$
and energy magnetization density $M^E$ for the dyonic black hole
are obtained by differentiating the free energy with respect to a
constant magnetic field for either the charge or momentum
currents,
\bea
M & = & - \frac{\delta \Omega}{\delta F_{xy}} \label{eq:m} \,, \\
M^E & = & - \frac{\delta \Omega}{\delta F^E_{xy}} \,.
\label{eq:me}
\eea
Here we define $\delta F^E_{xy} = \pa_x \delta g^0_{ty} - \pa_y
\delta g^0_{tx}$ and $\delta F_{xy} = \pa_x \delta A^0_y - \pa_y \delta A^0_x$, where $\delta
g^0_{ta}$ is a background gauge field sourcing $T_{ta}$, and
$\delta A_a^0$ sources $J_a$. Further comments on these magnetization
densities, and their computation for the scalar field theory in Eq.~(\ref{spsi})
appear in Appendix~\ref{app:mag}.

In the AdS/CFT correspondence, the free energy $\Omega$ is just
the action of the dual gravitational background. To compute the
derivatives in (\ref{eq:m}) and (\ref{eq:me}) we must consider on
shell fluctuations of the bulk metric and gauge field that tend
towards $\delta g^0_{ta}$ and $\delta A^0_{a}$, respectively, near
the conformal boundary $z \to 0$. We then differentiate the action
with respect to the boundary values of these fields.

More concretely, the boundary condition may be taken to be
\bea
\delta A_y & \to & x B \quad \text{as} \quad z \to 0 \,, \\
\delta g_{ty} \, z^2/\alpha \equiv \delta G_y  & \to & x B^E \quad \text{as} \quad z \to 0
\,,
\eea
with $B$ and $B^E$ constants and all other fields having
normalizable behavior near the boundary. It turns out that the
linearized fluctuation equations about the dyonic black hole
background with these boundary conditions may be consistently
truncated to the fields $\delta A_y$, $\delta A_t$ and $\delta
g_{ty}$. The solution is
\bea
\delta A_y & = & x (B - q B^E z) \,, \\
\delta G_y & = & x f(z) B^E \,, \\
\delta A_t & = & \frac{h B^E}{2 \a} (z^2-1) - \frac{h B}{q \a} (z-1)
\,.\label{eq:at}
\eea
Note that $A_t$ vanishes at the horizon, $z=1$, as required.

Because we are fluctuating about a solution, the linear variation
of the bulk action will vanish. However, there will be a boundary
term that arises due to integration by parts when evaluating the
action on the solution. There is also a boundary term that must be
included to renormalize the gravitational action
\be
I_\text{bdy.} = - \frac{\sqrt{2} N^{3/2}}{6 \pi} \left[\frac{1}{2}
\int d^3 x \sqrt{-\gamma}\,\theta +
\int d^3 x \sqrt{- \gamma} \right] \,,
\label{eq:I-ren}
\ee
where $\gamma$ is the boundary metric and
$\theta=\gamma^{\mu\nu}\theta_{\mu\nu}$ is the trace of the
extrinsic curvature $\theta_{\mu \nu} = - \frac12(\nabla_{\mu}
n_{\nu}+ \nabla_\nu n_\mu)$, with $n$ an outward directed unit
normal vector to the boundary.

For fluctuations about a solution, we have
\be\label{eq:varS}
\delta I = \frac{\sqrt{2} N^{3/2}}{6 \pi} \int d^3x \sqrt{-\gamma}
\left[- F_{\mu \nu} n^{\nu} \delta A^{\mu} + \frac{1}{4} \left(\theta^{\mu \nu}
- \theta \gamma^{\mu \nu} - 2 \gamma^{\mu \nu} \right)
\delta g_{\mu \nu}\right] \,.
\ee
Evaluated on the background (\ref{eq:metric}) and
(\ref{eq:Ffield}), and considering only $\delta A_y$, $\delta A_t$
and $\delta g_{ty}$, this expression becomes
\be
\delta I = \frac{\sqrt{2} N^{3/2}}{6 \pi} q \a^2 \int d^3x \delta A_t \,.
\ee
Note that only the first term in the variation of the action
(\ref{eq:varS}) contributes. Using the solution (\ref{eq:at}) for
$\delta A_t$, we obtain the magnetization
\be
M = - \frac{\delta S}{\delta B} = - \frac{\sqrt{2} N^{3/2}}{6 \pi}
\a h = - \frac{\sqrt{2} N^{3/2}}{6 \pi} \frac{B}{\a} \,,
\ee
in complete agreement with our previous expression (\ref{eq:mag}).
For the energy magnetization we obtain
\be
M^E = - \a \frac{\delta S}{\delta B^E} = - \frac{q \a}{2} M =
\frac{\mu \, M}{2} \,.
\ee

\subsection{Transport coefficients in the d.c. limit}

We will obtain the transport coefficients using Kubo formulae for
the retarded Greens functions. The Greens functions are obtained
by considering fluctuations about the dyonic black hole
background. In \cite{hartnoll} these functions were obtained at
$k=0$ and to leading order as $\w \to 0$ with $B$ and $\rho$ held
fixed. Unlike in our general magnetohydrodynamic (MHD) analysis in Section~\ref{sec:mhd}, no
assumptions are made here requiring $B$ to be small.
The
current-current correlator is
\be
G^R_{J^i J^j}(\w) = - \frac{\rho}{B} i \w \ep_{ij} \,,
\ee
the current-momentum correlator is
\be\label{eq:jt}
G^R_{J^i T^{tj} }(\w) = - \frac{3 \vep}{2 B} i \w \ep_{ij} \,,
\ee
and the momentum-momentum correlator is
\be\label{eq:tt}
G^R_{T^{ti} T^{tj} }(\w) = - \frac{\chi T^3 s^2}{\chi^2 B^2 +
\rho^2 T^2} i \w \delta_{ij} -
\frac{9 \, \rho \, \vep^2 T^2}{4 B (\chi^2 B^2+\rho^2 T^2)} i \w \ep_{ij}
\,.
\ee
To relate these results to our general MHD study, we need an expression for the
heat current $Q^i$. This is defined in Eq.~(\ref{heat}); using the expression for the
stress energy tensor in Eq.~(\ref{perturbs1}), we see that in linear response (small velocities with respect to the lab frame) we can work with
\be
Q^i = T^{ti} - \mu J^i \, ,
\ee
leading to the correlators
\be
G^R_{Q^i J^{j}}(\w) = \left(\frac{- s T}{B} + M \right) i \w
\ep_{ij}
\,,
\ee
and
\be
G^R_{Q^i Q^{j}}(\w) =  - \frac{\chi T^3 s^2}{\chi^2 B^2 + \rho^2
T^2} i \w \delta_{ij} + \frac{- \rho s^2 T^4 + B^2
\mu^2 \rho \chi^2 + \rho T^2 M^2 B^2}{B(\chi^2 B^2 +
\rho^2 T^2)} i \w \ep_{ij} \,.
\ee
In obtaining these expressions, we used (\ref{eq:rel1}) and
(\ref{eq:rel2}).

The electrical conductivity is given by the Kubo formula
\be\label{eq:stringsigma}
\sigma_{ij} = - \lim_{\w \to 0} \frac{\textrm{Im} G_{J^i J^j}^R(\w)}{\w} =\frac{\rho}{B} \epsilon_{ij}  \,.
\ee

The other thermoelectric tensors are also given by a Kubo formula.
However, we should use the transport currents which are obtained from the supergravity currents by subtracting the magnetization currents~\cite{cooper}. The correct Kubo formula is
\be
\alpha_{ij} = - \frac{1}{T} \lim_{\w \to 0} \frac{\textrm{Im}
 G_{J^i Q^j}^R(\w)}{\w} + \frac{M}{T} \epsilon_{ij} \,.
\ee
Thus we obtain
\be\label{eq:stringalpha}
\alpha_{ij} = \frac{s}{B} \ep_{ij}  \,.
\ee
Similarly, the heat conductivity is given by the Kubo formula
\be
\overline{\kappa}_{ij} = - \frac{1}{T} \lim_{\w \to 0} \frac{\textrm{Im}
 G_{Q^i Q^j}^R(\w)}{\w} + \frac{2 (M^E - \mu M)}{T} \epsilon_{ij} \,,
\ee
or
\be\label{eq:stringkappa}
 \overline{\kappa}_{ij} = \frac{\chi T^2 s^2}{\chi^2
B^2 + \rho^2 T^2} \delta_{ij} + \frac{\rho s^2 T^4}{T B(\chi^2 B^2
+ \rho^2 T^2)} \ep_{ij}\,.
\ee

We can now compare these results with those of our general MHD
computations, by taking the $\w \to 0$ limit of the MHD transport
coefficients (\ref{fres1}). We see immediately that the
expressions for $\sigma_{ij}$ and $\alpha_{ij}$ agree exactly. In
order to match $\overline{\kappa}_{ij}$ we need to recall that the
MHD results are only valid for small magnetic fields $B
\ll T^2$, see Eq.~(\ref{small}). Furthermore, we need to know the
conductivity $\sigma_Q$ for the dyonic black hole. It is shown in
Ref.~\onlinecite{hh} that for the black hole
\be
\left. \sigma_Q \right|_{B=0} = \left( \frac{T s}{\vep + P} \right)^2 \frac{\chi}{T} \,.
\ee
Using this formula and taking the small magnetic field limit, we obtain
\be
\left. \overline{\kappa}_{ij} \right|_{B \ll T^2}  = \frac{\sigma_Q (\vep+P)^2}{\rho^2 T} \delta_{ij} + \frac{s^2 T}{B \rho} \ep_{ij}\,.
\ee
This expression now agrees exactly with the corresponding limit of
the MHD result (\ref{fres1}). Thus we see that the dyonic black hole fits into the general
class of finite temperature deformations of quantum critical
points that we have studied via hydrodynamics. The black hole gives
specific values for $\sigma_Q$ and the other thermodynamic quantities and
furthermore allows the results to be extended to arbitrary magnetic field.

\subsection{Bulk electromagnetic duality and CFT particle/vortex duality}

A consequence of the bulk description is that it gives a very
transparent rationale behind the dualities in the
transport coefficients that we commented upon earlier. The study
of electromagnetic duality in the dyonic black hole is pursued in depth in Ref.~\onlinecite{hh}, which
furthermore obtains expressions for the black hole transport coefficients
away from the d.c. limit. Here we shall summarize some results from that paper
and show how they precisely match our expectations from MHD.

The bulk Maxwell theory enjoys an electromagnetic duality. This
interchanges the bulk electric and magnetic fields
${\boldsymbol{E}}
\to {\boldsymbol{B}}$ and ${\boldsymbol{B}} \to - {\boldsymbol{E}}$. Acting
on the dyonic black hole solutions (\ref{eq:metric}), this
corresponds to $q \to h$ and $h \to -q$. Using (\ref{eq:charges})
and (\ref{eq:chargedensity}) and the fact that the bulk coupling
is inverted under electromagnetic duality, this implies that
\be\label{eq:adstrans}
\quad B \to \frac{ T \rho}{\chi} \,, \quad \rho \to - \frac{
T B}{\chi} \,, \quad \frac{\chi}{T} \to \frac{T}{\chi} \,.
\ee
Thus we see that the bulk electromagnetic duality maps the CFT
into the same CFT with the values of the background magnetic field
and charge density interchanged. This is the origin of the
particle-vortex duality that we noted in our MHD computations.
Indeed it is immediately seen that under the transformation
(\ref{eq:adstrans}) our expressions for $\sigma, \alpha$ and $\bar
\kappa$ in Eqs. (\ref{eq:stringsigma}), (\ref{eq:stringalpha}) and
(\ref{eq:stringkappa}) transform according to (\ref{dualtrans}).
There are some overall factors of $\chi/T$ different to
(\ref{dualtrans}), due to the fact that the transformation
(\ref{eq:adstrans}) is normalized differently to
(\ref{eq:exchange}). The normalization in (\ref{eq:adstrans}) is
natural from the string perspective.

It remains to see how the thermoelectric transport coefficients of
the black hole transform under this map. This is shown in detail
in Ref.~\onlinecite{hh}. The central point is the following. The
bulk Maxwell potential $A$ determines the bulk electric and
magnetic fields through ${\boldsymbol{E}} \sim \pa_t A$ and
${\boldsymbol{B}} \sim
\pa_z A$. As we tend towards the boundary $z \to 0$, the electric
piece is non-normalizable, and results in a boundary electric
field $E$. The magnetic piece however is normalizable and
therefore results in a boundary current $J$. The bulk
electromagnetic duality is thus seen to exchange the electric
field in the CFT with the current. More precisely, one finds
\be
E_i \leftrightarrow \epsilon_{ij} J_j \,.
\ee
As we commented below Eq.~(\ref{dualtrans}), this map together
with the definition of the transport coefficients in
Eq.~(\ref{alltrans}) is enough to obtain all the duality
transformations (\ref{dualtrans}).

The results for the transport coefficients presented in the
previous subsection were only obtained in the d.c. limit $\w \to
0$ with $B$ and $\rho$ fixed, following
Ref.~\onlinecite{hartnoll}. However, using the AdS/CFT dictionary,
it is possible to study thermoelectric transport at all
frequencies. This is done in Ref.~\onlinecite{hh}. In particular,
taking the limit $\w \to 0$ with $\rho \sim B \sim T^{3/2}
\sqrt{\w}$, one obtains precisely the same expressions as those
that followed from the MHD analysis (\ref{fres3}), thus providing
a consistency check for our picture. One can go further with the
dyonic black hole and study transport and all $\rho,B$ and $\w$
numerically. For all values, the particle-vortex duality holds
automatically because of the bulk electromagnetic duality. This is
the power of the AdS/CFT correspondence: all transport phenomena
of the strongly correlated CFT at large $N$ are reduced to solving
the equations for classical perturbations of the dual black hole
in Einstein-Maxwell theory.

\section{Conclusions}
\label{sec:conc}

This paper has presented a general theory for hydrodynamic thermal and electric transport in
in the vicinity of a quantum critical points described by
``relativistic'' quantum field theories. We have also shown that the results
constitute a valuable starting point to understand experimental observations in a regime
where no previous description was available.

It is perhaps useful to describe the results here from
the vantage point of the Galilean-invariant hydrodynamic approaches which are traditionally
used in condensed matter physics \cite{km,pethick}. In such theories, charge (or number)
currents are proportional to the momentum current, and consequently the conductivity is infinite in the absence
of impurities
(in the presence of a magnetic field, this implies Kohn's theorem \cite{kohn}). The natural transport
co-efficient is the thermal conductivity, and this determines various diffusivities and damping constants.

In contrast, in the present paper, we have used a very different starting point. We considered a theory
with both particle and anti-particle (hole) excitations, in which there is no proportionality between
momentum and charge currents. For the case with particle-hole symmetry ($\rho=0$) and
$B=0$, the momentum and charge currents are decoupled from each other, and it is possible to have
a charge current with no momentum current: the electrical current can decay to
zero from such a state, and this decay is associated with the universal electrical conductivity
$\sigma_Q$ \cite{damle}. There is no analog to $\sigma_Q$ in the Galilean invariant case. Also, because
of the symmetry of the stress-energy tensor, we could identify the energy current with the
momentum density; the conservation of momentum then implied that the thermal
conductivity was infinite \cite{vojta}.
Upon relaxing the requirement of particle-hole symmetry ({\em i.e.\/} allowing $\rho \neq 0$), we found
the appearance of some characteristics of the Galilean invariant systems. In Eq.~(\ref{ds1})
we found that the excess particles (or holes) contributed a Drude-like conductivity
above the quantum-critical $\sigma_Q$. The thermal conductivity became finite,
but with a value related to $\sigma_Q$ by a Wiedemann-Franz-like relation. Finally, we also
turned on a $B \neq 0$, and showed how all of the longitudinal and transverse
transport co-efficients could be related to $\sigma_Q$ in relationships that were
summarized in Section~\ref{sec:intro}.

While our analysis was specialized to relativistic quantum critical points, we expect that
many aspects will generalize to other strongly interacting quantum critical points. Only a discrete
particle-hole symmetry is required to decouple the charge and momentum currents, and this should
be sufficient to obtain a finite $\sigma_Q$.

We also discussed applications of this general hydrodynamic structure to measurements of
the Nernst co-efficient in the cuprates and NbSi films. For reasonable values of the parameters,
we were able to reproduce several key aspects of the $B$ and $T$ dependence of the observations.
Our results also make a significant prediction, characteristic of such ``relativistic'' theories of the
superfluid-insulator transition: the presence of a hydrodynamic cyclotron mode.
For the simplest case of a superfluid-insulator transition of Cooper pairs at integer
filling as described by Eq.~(\ref{spsi}), this cyclotron mode can be considered due to the motion
of Cooper pairs (or their dual vortices). However, for the more complicated examples at fractional
filling noted in Section~\ref{sec:systems}, such a simple interpretation is not possible, and the cyclotron
mode is due to motion of all charge carriers, including those carrying fractional charges.
From our fits to the data in the cuprates in Section~\ref{sec:expts},
we found that in presently studied samples this cyclotron mode was strongly overdamped by impurity
scattering. However, this raises the possibility that
the cyclotron resonance which might be observable in ultrapure samples. We estimated that
the hydrodynamic cyclotron frequency in the cuprates
was smaller than the cyclotron frequency of free electrons by a factor of order $10^{-3}$.
Observation of this resonance would constitute a strong test of the theoretical ideas presented here, and
we hope such experiments will be undertaken.

Another class of results in this paper described the remarkable holographic connection between this
intricate hydrodynamic behavior in 2+1 dimensions and the quantum theory of dyonic black holes in 3+1 dimensions.
Using the AdS/CFT
connection, we presented in Section~\ref{sec:dyon} exact results for the hydrodynamic response
functions of the vicinity of a 2+1 dimensional supersymmetric conformal field theory. In the appropriate limiting
regime, these results, and those in Ref.~\onlinecite{hh}, were in complete agreement with those obtained from  the magnetohydrodynamic
analysis in Section~\ref{sec:mhd}. This agreement lends strong support to
the validity of our MHD analysis. Indeed, the analysis of the dual gravity
theory helped guide our determination of the MHD equations.

\acknowledgments
Related results are reported in a companion paper by C.~Herzog and one of us \cite{hh},
and we thank C.~Herzog for numerous valuable discussions.
We acknowlege A.~Vishwanath for pointing out that Eq.~(\ref{T4}) agrees with the $T$ dependence of recent experimental data.
We also thank P.~W.~Anderson, H.~Aubin, M.~J.~Bhaseen, K.~Behnia,
A.~G.~Green, D.~Podolsky, N.~P.~Ong, S.~Shapiro, D.~T.~Son,
S.~L.~Sondhi, A.~Strominger, and A.~Vishwanath for useful
discussions. This research was supported by the NSF under grants
PHY05-51164 (PK and SH) and DMR-0537077 (MM and SS) and by the
Swiss National Fund for Scientific Research under grant
PA002-113151 (MM).

\appendix

\section{Normal modes of the hydrodynamic equations}
\label{app:modes}

It is interesting to analyze the normal modes of the linearized magnetohydrodynamic equations in the absence of impurity scattering. Assuming a space and time dependence $e^{-i\w t + i\vec{k} \cdot \vec{x}}$ of $\delta\mu,\,\delta T$ and $v_\parallel := \vec{v}\cdot \vec{k}/k$ and $v_\perp := (\vec{k}/k)\cdot (\hat{\epsilon}\vec{v})$, we find four independent modes satisfying the equations
\bea
\w \left(\left. \frac{\partial \epsilon}{\partial \mu} \right|_{T} \del \mu + \left. \frac{\partial \epsilon}{\partial T} \right|_{\mu} \del T\right) - k (\ep+P)v_{\parallel} & = & 0 \,, \nonumber \\
\w \left(\left. \frac{\partial \rho}{\partial \mu} \right|_{T} \del \mu + \left. \frac{\partial \rho}{\partial T} \right|_{\mu} \del T\right) - k \left(\rho v_\parallel + B \sigma_Q v_{\perp} \right) + i k^2 \sigma_Q \left( \del \mu - \frac{\mu}{T} \del T\right) & = & 0 \,, \nonumber \\
\w (\ep + P) v_{\parallel} - k (\rho \del \mu + s \del T) - i \rho B v_\perp + i B^2 \sigma_Q v_{\parallel} - i k^2 (\eta + \zeta) v_{\parallel} & = & 0 \,, \nonumber \\
\w (\ep + P) v_{\perp} + k B \sigma_Q \left( \del \mu - \frac{\mu}{T} \del T\right) + i \rho B v_\parallel + i B^2 \sigma_Q v_\perp - i k^2 \eta v_\perp & = & 0 \,.
\eea
In the long wavelength limit $k \ll 1$, one finds
two modes corresponding to damped cyclotron oscillations of the plasma
\be
\w_\pm = \pm \omega_c - i\gamma \,.
\ee
These modes have a velocity field with $v_\parallel = \pm i v_\perp$, while $\del \mu$ and $\del T$ are smaller than $v_\parallel, v_\perp$ by a factor of order ${\mathcal{O}}(k)$.

Further there is a diffusive mode with frequency proportional to the conductivity $\sigma_Q$  and a quadratic dispersion relation
\be
\w_{\rm diff} = -i\frac{ k^2 \sigma_Q (\ep + P)^2}{\displaystyle  T \left(\left. \frac{\partial \epsilon}{\partial \mu} \right|_{T} \left. \frac{\partial \rho}{\partial T} \right|_{\mu}-\left. \frac{\partial \epsilon}{\partial T} \right|_{\mu} \left. \frac{\partial \rho}{\partial \mu} \right|_{T}\right) (\rho^2 + B^2 \sigma_Q^2)} \,.
\ee
This mode has no fluctuations in energy density, $\del \ep =\displaystyle  \left. \frac{\partial \ep}{\partial \mu} \right|_{T} \del \mu +\left. \frac{\partial \ep}{\partial T} \right|_{\mu} \del T=0$. The velocity field  $\vec{v}$ is of order ${\mathcal{O}}(k)$ relative to $\del \mu, \del T$.

Finally there is a subdiffusive, transverse shear mode with strongly suppressed fluctuations in temperature and longitudinal velocity component $\del T={\mathcal O}(k^2)$, $v_{\parallel} = {\mathcal O}(k^3)$. It exhibits an unusual dispersion relation
\be
\label{subdiff}
\w_{\rm subdiff} = \frac{i k^4 \eta }{\displaystyle B^2 \left. \frac{\partial \rho}{\partial \mu} \right|_{T}} \,,
\ee
and we have the relation
\be
\label{relationmode4}
i k \delta\mu \approx B v_\perp.
\ee

The origin of the $k^4$ dispersion (\ref{subdiff}) can be seen as follows: A strongly suppressed
$\omega(k)$ implies that momentum density is nearly conserved. Hence the total force density vanishes to lowest order, i.e., the Lorentz force is balanced by a longitudinal pressure gradient and a transverse friction force,
\bea
(\vec{J}\times\vec{B})_\parallel &\approx& \vec{\nabla}P +{\mathcal O}(\eta,\zeta,\omega),\\
\eta\nabla^2 \vec{v}_\perp  & \approx & (\vec{J}\times\vec{B})_\perp=-BJ_\parallel.
\eea
The first equation yields relation (\ref{relationmode4}). The second can be injected into the equation for charge conservation
\be
\partial_t \rho =  -\vec{\nabla}\cdot\vec{J}\approx \vec{\nabla}\cdot\left[\frac{\eta}{B} \nabla^2 (v_\perp)\right]\approx \vec{\nabla}\cdot\left[\frac{\eta}{B} \nabla^2 \left(\frac{\vec{\nabla}\mu}{B}\right)\right],
\ee
from which the dispersion follows upon using $\partial_t \rho= \displaystyle -i\omega \left. \frac{\partial \rho}{\partial \mu} \right|_{T} \del \mu$.

\section{Magnetization and energy magnetization}
\label{app:mag}

In our computation of the transport co-efficients using the Kubo formula in Section~\ref{sec:dyon},
we had to face the issue of the subtraction of magnetization currents, as discussed earlier in
Refs.~\onlinecite{streda,cooper}. These subtractions were computed in Section~\ref{sec:dyon}
using the AdS/CFT mapping. This appendix describes the nature of these magnetization subtractions
in the context of the scalar field theory in Eq.~(\ref{spsi}). Actually, most of the basic issues are
already clarified in free field theory, and so we will limit our discussion here to this simple case.
The generalization of the free field results to the interacting Wilson-Fisher fixed points can be
straightforwardly carried out along the lines of Refs.~\onlinecite{csy,zphys}, and so we will
not discuss it here.

So we consider here the free field version of Eq.~\ref{spsi} with Lagrangian
\begin{equation}
\mathcal{L} =  \left[( \partial^\mu + i A^\mu) \psi^\ast \right]
\left[(
\partial_\mu - i A_\mu ) \psi \right] + m_0^2 |\psi|^2.
\end{equation}
The stress-energy tensor is \cite{hawking}
\begin{equation}
T_{\mu\nu} = \left[( \partial_\mu + i A_\mu) \psi^\ast \right]
\left[(
\partial_\nu - i A_\nu ) \psi \right] + \left[( \partial_\nu + i A_\nu) \psi^\ast \right]
\left[(
\partial_\mu - i A_\mu ) \psi \right] - g_{\mu\nu} \mathcal{L},
\label{tmn}
\end{equation}
while the U(1) current is
\begin{equation}
J_\mu = i \psi^\ast ( \partial_\mu - i A_\mu) \psi - i \psi (
\partial_\mu + i A_\mu) \psi^\ast .
\end{equation}
The equation of motion is
\begin{equation}
( \partial_\mu - i A_\mu) (\partial^\mu - i A^\mu ) \psi = m_0^2 \psi
.
\end{equation}
It is now a straightforward, but tedious, exercise to verify that
the above expressions do indeed imply the MHD equation of motion
in Eqs.~(\ref{e1p},\ref{e1}).

For the thermodynamics, we need the particle and hole eigenenergies.
These are organized in Landau levels, with energy
\begin{equation}
\epsilon_\ell^2 = 2B (\ell+1/2) + m_0^2,
\end{equation}
($\ell = 0, 1, \ldots \infty$) and degeneracy per unit area of $B/(2
\pi)$. From this, we can easily obtain expressions for the grand
potential (for all thermodynamic quantities we subtract an infinite
$T=0$ value, and $\omega_n$ is a Matsubara frequency which is an integer
multiple of $2 \pi$):
\begin{eqnarray}
\frac{\Omega}{\mathcal{V}} = -P &=& \frac{BT}{2 \pi} \sum_{\omega_n} \sum_\ell \ln
\left[ (\omega_n - i \mu)^2 + \epsilon_\ell^2 \right]
\nonumber \\
&=& \frac{BT}{2 \pi} \sum_{\ell} \left[ \ln \left( 1 -
e^{-(\epsilon_\ell - \mu)/T} \right) +\ln \left( 1 -
e^{-(\epsilon_\ell + \mu)/T} \right) \right]. \label{omega}
\end{eqnarray}
We also obtain the entropy, $s$, the density, $\rho$, and the
magnetization density, $M$, by
\begin{equation}
s = - \frac{1}{\mathcal{V}}\frac{\partial \Omega}{\partial T}~~;~~ \rho = - \frac{1}{\mathcal{V}}\frac{\partial
\Omega}{\partial \mu}~~;~~M=-  \frac{1}{\mathcal{V}}\frac{\partial \Omega}{\partial B}\,.
\label{pm}
\end{equation}
Following Cooper {\em et al.} \cite{cooper}, it is useful to define an internal
pressure $P_{\rm int}$ which equals
\begin{eqnarray}
P_{\rm int} &=& P - M B \nonumber \\
&=& \frac{B^2}{2\pi} \sum_{\ell} \frac{\ell+1/2}{\epsilon_\ell}
\left[ \frac{1}{e^{(\epsilon_\ell - \mu)/T} - 1} +
\frac{1}{e^{(\epsilon_\ell + \mu)/T} - 1} \right].
\end{eqnarray}

We define the energy density $\varepsilon$ by $\langle T_{tt} \rangle$.
Evaluating this from (\ref{tmn}) in Euclidean Matsubara space, we
obtain
\begin{eqnarray}
\varepsilon = \langle T_{tt} \rangle &=& \frac{B}{2\pi} \sum_{\ell}
T\sum_{\omega_n}
 \frac{- (\omega_n-i \mu)^2 +
\epsilon_\ell^2}{(\omega_n-i \mu)^2 + \epsilon_\ell^2} \nonumber
\\
&=& \frac{B}{2\pi} \sum_{\ell} \epsilon_\ell \left[
\frac{1}{e^{(\epsilon_\ell - \mu)/T} - 1} +
\frac{1}{e^{(\epsilon_\ell + \mu)/T} - 1} \right].
\end{eqnarray}
It is now easily verified that the relations in Eq.~(\ref{thermo}) are
obeyed.

In a similar manner, we can compute $\langle T_{xx} \rangle$ and
find
\begin{eqnarray}
\langle T_{xx} \rangle &=& \frac{B}{2\pi} \sum_{\ell}
T\sum_{\omega_n}
 \frac{- (\omega_n-i \mu)^2 -m_0^2}{(\omega_n-i \mu)^2 + \epsilon_\ell^2} \nonumber
\\
&=& \frac{B^2}{2\pi} \sum_{\ell} \frac{\ell+1/2}{\epsilon_\ell}
\left[ \frac{1}{e^{(\epsilon_\ell - \mu)/T} - 1} +
\frac{1}{e^{(\epsilon_\ell + \mu)/T} - 1} \right] \nonumber \\
&=& P_{\rm int},
\end{eqnarray}
and so Eqs.~(\ref{tpint}) are also obeyed.

Let us now write down the explicit result for $M$ from
Eq.~(\ref{pm}):
\begin{eqnarray}
M &=& - \frac{T}{2 \pi} \sum_\ell \left[ \ln \left( 1 -
e^{-(\epsilon_\ell - \mu)/T} \right) +\ln \left( 1 -
e^{-(\epsilon_\ell + \mu)/T} \right) \right] \nonumber \\
&~&~~~~-\frac{B}{2\pi} \sum_{\ell} \frac{\ell+1/2}{\epsilon_\ell}
\left[ \frac{1}{e^{(\epsilon_\ell - \mu)/T} - 1} +
\frac{1}{e^{(\epsilon_\ell + \mu)/T} - 1} \right].
\end{eqnarray}
A direct evaluation of $\langle \vec{r} \times \vec{J} \rangle$ in
an infinite system yields only the second term but not the first. We
now argue that the first term is the contribution of edge states.
Notice that this first term can be rewritten in the form
\begin{eqnarray}
M &=&  \frac{1}{2 \pi} \int_0^{\infty} dE g_{e} (E) \left[
\frac{1}{e^{(E - \mu)/T} - 1} +
\frac{1}{e^{(E + \mu)/T} - 1} \right]\nonumber \\
&~&~~~~-\frac{B}{2\pi} \sum_{\ell} \frac{\ell+1/2}{\epsilon_\ell}
\left[ \frac{1}{e^{(\epsilon_\ell - \mu)/T} - 1} +
\frac{1}{e^{(\epsilon_\ell + \mu)/T} - 1} \right],
\end{eqnarray}
where we can interpret $g_{e} (E)$ as the magnetization of edge
states:
\begin{equation}
g_e (E) = \left\{
\begin{array}{ccl} 0 & ~,~ & \mbox{for $E< \epsilon_0$} \\
\ell & ~,~ & \mbox{for $\epsilon_{\ell-1} < E < \epsilon_{\ell}$}
 \end{array}
\right. .
\end{equation}
Note that $g_e (E)$ is a piecewise constant function, and there is
one additional edge state as each Landau level is crossed, as
expected. We can write the expression for the magnetization
as a sum of a bulk and edge contributions as
\begin{equation}
M =  \frac{1}{2 \pi} \int_0^{\infty} dE g (E) \left[ \frac{1}{e^{(E
- \mu)/T} - 1} + \frac{1}{e^{(E + \mu)/T} - 1} \right],
\end{equation}
where
\begin{equation}
g(E) = g_e (E) - \sum_{\ell=0}^{\infty}
\frac{B(\ell+1/2)}{\epsilon_\ell} \delta(E- \epsilon_\ell).
\end{equation}

With the above form for $M$, we can now immediately use the results
of Ref.~\onlinecite{streda} (compare their
Eqns. (31) and (34)) to obtain the value of $M^E$:
\begin{equation}
M^E =  \frac{1}{2 \pi} \int_0^{\infty} dE E g (E) \left[
\frac{1}{e^{(E - \mu)/T} - 1} - \frac{1}{e^{(E + \mu)/T} - 1}
\right].
\end{equation}
The subtraction for $\overline{\kappa}_{xy}$ is, by Cooper {\em et al.} \cite{cooper} Eq.
(69), $2M^Q$ where
\begin{eqnarray}
M^Q &=& M^E - \mu M \nonumber \\
&=& \frac{1}{2 \pi} \sum_{\ell} \left[ \int_{\epsilon_\ell -
\mu}^{\epsilon_\ell + \mu} \frac{E dE}{e^{E/T} - 1}  - \frac{B (\ell
+ 1/2)}{\epsilon_\ell} \left(\frac{\epsilon_\ell -
\mu}{e^{(\epsilon_\ell - \mu)/T} - 1} - \frac{\epsilon_\ell +
\mu}{e^{(\epsilon_\ell + \mu)/T} - 1}\right)\right].
\label{emres1}
\end{eqnarray}

An alternative evaluation of $M^Q$, which is generalizable to
interacting theories, follows from the representation of the heat
current discussed in Ref.~\onlinecite{moreno}. As noted in Section~\ref{sec:dyonmag},
we need the response
to a ``magnetic field'' which associated with the energy (or heat) current, just
as the ordinary magnetization is a response to a magnetic field associated
with the charge current. So we introduce
a vector potential $\vec{A}_Q$ which couples to heat current: this is done
by the replacement \cite{moreno} $\vec{A} \rightarrow \vec{A} + i \vec{A}_Q \omega_n$.
Consequently, the free energy density in the presence of this additional
``magnetic field'' $B_Q$ is
 obtained from
Eq.~(\ref{omega}) simply by the replacement $B \rightarrow B + i B_Q
\omega_n$, which yields
\begin{equation}
\frac{\Omega}{\mathcal{V}} = \frac{T}{2 \pi} \sum_\ell \sum_{\omega_n} (B + i B_Q
\omega_n) \ln \left[ (\omega_n - i \mu)^2 + 2 (B+i B_Q
\omega_n)(\ell + 1/2) + m_0^2 \right].
\end{equation}
Taking the derivative with respect to $B_Q$ we obtain
\begin{eqnarray}
M^Q &=& - \frac{1}{\mathcal{V}}\frac{\partial \Omega}{\partial B_Q} \nonumber \\
&=& \frac{T}{2 \pi} \sum_\ell \sum_{\omega_n} \left( -i \omega_n \ln
\left[ (\omega_n - i \mu)^2 + \epsilon_\ell^2 \right] + \frac{-2 i
\omega_n B(\ell + 1/2)}{(\omega_n - i \mu)^2 + \epsilon_\ell^2}
\right) \nonumber \\
&=& \frac{T}{2 \pi} \sum_\ell \sum_{\omega_n} \left( -i \omega_n \ln
\left[ -i \omega_n + \epsilon_\ell - \mu \right] -i \omega_n \ln
\left[ i \omega_n + \epsilon_\ell + \mu \right]\right)\nonumber
\\ &~&~~~~~+\frac{1}{2 \pi} \sum_\ell \left [  - \frac{B (\ell +
1/2)}{\epsilon_\ell} \left(\frac{\epsilon_\ell -
\mu}{e^{(\epsilon_\ell - \mu)/T} - 1} - \frac{\epsilon_\ell +
\mu}{e^{(\epsilon_\ell + \mu)/T} - 1}\right) \right] \nonumber \\
&=& \frac{1}{2 \pi} \sum_{\ell} \left[ \int_{\epsilon_\ell -
\mu}^{\epsilon_\ell + \mu} \frac{E dE}{e^{E/T} - 1}  - \frac{B (\ell
+ 1/2)}{\epsilon_\ell} \left(\frac{\epsilon_\ell -
\mu}{e^{(\epsilon_\ell - \mu)/T} - 1} - \frac{\epsilon_\ell +
\mu}{e^{(\epsilon_\ell + \mu)/T} - 1}\right)\right],
\end{eqnarray}
which agrees with Eq.~(\ref{emres1}).


\begin{thebibliography}{99}

\bibitem{ong1} Z.~A.~Xu, N.~P.~Ong, Y.~Wang, T.~Kakeshita, and
S.~Uchida, Nature {\bf 406}, 486 (2000).

\bibitem{ong2}  Y.~Wang, L.~Li, M.~J.~Naughton, G.~D.~Gu, S.~Uchida,
and N.~P.~Ong, Phys. Rev. Lett. {\bf 95}, 247002 (2005).

\bibitem{ong3} L.~Li, J.~G.~Checkelsky, S.~Komiya, Y.~Ando, N.~P.~Ong,
Nature Physics {\bf 3}, 311 (2007).

\bibitem{ong4}  Y.~Wang, L.~Li,
and N.~P.~Ong, Phys. Rev. B {\bf 73}, 024510 (2006).

\bibitem{behnia1} A.~Pourret, H.~Aubin, J.~Lesueur, C.~A.~Marrache-Kikuchi, L.~Berg\'e,
L.~Dumoulin, and K.~Behnia, Nature Physics {\bf 2}, 683 (2006).

\bibitem{behnia2} A.~Pourret, H.~Aubin, J.~Lesueur, C.~A.~Marrache-Kikuchi, L.~Berg\'e,
L.~Dumoulin, and K.~Behnia, arXiv:cond-mat/0701376.

\bibitem{vadim} V.~Oganesyan and I.~Ussishkin, Phys. Rev. B {\bf 70}, 054503 (2004).

\bibitem{ussishkin} I.~Ussishkin, S.~L.~Sondhi, and D.~A.~Huse,
Phys.~Rev.~Lett. {\bf 89}, 287001 (2002).

\bibitem{mukerjee} S.~Mukerjee and D.~A.~Huse, Phys. Rev. B {\bf 70}, 014506 (2004).

\bibitem{podolsky} D.~Podolsky, S.~Raghu, and A.~Vishwanath, arXiv:cond-mat/0612096.

\bibitem{pwa} P.~W.~Anderson, arXiv:0705.1174.

\bibitem{bgs}  M.~J.~Bhaseen, A.~G.~Green, and S.~L.~Sondhi,
Phys. Rev. Lett. {\bf 98}, 166801 (2007).

\bibitem{ssbook} S.~Sachdev, {\em Quantum Phase Transitions},
Cambridge University Press, Cambridge (1999).

\bibitem{m2cft} C.~P.~Herzog, P.~Kovtun. S.~Sachdev,
and D.~T.~Son, Phys. Rev. D {\bf 75}, 085020 (2007).

\bibitem{mpaf} M.~P.~A.~Fisher, G.~Grinstein, and S.~M.~Girvin,
Phys.\ Rev.\ Lett.\ {\bf 64}, 587 (1990).

\bibitem{wenzee} X.-G.~Wen and A.~Zee,  Int. J. Mod. Phys. B {\bf 4},
437 (1990).

\bibitem{damle} K.~Damle and S.~Sachdev, Phys.\ Rev.\ B {\bf 56}, 8714 (1997).

\bibitem{troyer} M.~Troyer and S.~Sachdev, Phys. Rev. Lett. {\bf 81}, 5418 (1998).

\bibitem{balents1} L.~Balents, L.~Bartosch, A.~Burkov, S.~Sachdev,
and K.~Sengupta,  Phys. Rev. B {\bf 71}, 144508 (2005).

\bibitem{lorenz} L.~Balents, L.~Bartosch, and S.~Sachdev,
Annals of Physics {\bf 321}, 1528 (2006).

\bibitem{fwgf} M.~P.~A.~Fisher, P.~B.~Weichman, G.~Grinstein, and D.~S.~Fisher,
Phys.\ Rev.\ B {\bf 40}, 546 (1989).

\bibitem{peskin} M.~E.~Peskin, Annals of Physics, {\bf 113}, 122 (1978).

\bibitem{dasgupta} C.~Dasgupta and B.~I.~Halperin, Phys.\ Rev.\ Lett.\ {\bf 47},
1556 (1981).

\bibitem{senthil} T.~Senthil, A.~Vishwanath, L.~Balents, S.~Sachdev, and
M.~P.~A.~Fisher,  Science
{\bf 303}, 1490 (2004)

\bibitem{balents2} L.~Balents, L.~Bartosch, A.~Burkov, S.~Sachdev,
and K.~Sengupta,  Phys. Rev. B {\bf 71}, 144509 (2005).

\bibitem{courtney} C.~Lannert, M.~P.~A.~Fisher, and T.~Senthil, Phys.
Rev. B {\bf 63}, 134510 (2001).

\bibitem{balents3} L.~Balents and S.~Sachdev, Annals of Physics, in press, arXiv:cond-mat/0612220.

\bibitem{herzog}  C.~P.~Herzog,  JHEP {\bf 0212}, 026 (2002).

\bibitem{hartnoll} S.~Hartnoll and P.~Kovtun, arXiv:0704.1160.

\bibitem{hh} S.~Hartnoll and C.~P.~Herzog, arXiv:0706.3228.

\bibitem{csy} A.~V.~Chubukov, S.~Sachdev, and J.~Ye, Phys. Rev. B
{\bf 49}, 11919 (1994); S.~Sachdev, Phys. Lett. B {\bf 309}, 285 (1993).

\bibitem{zphys} S.~Sachdev, Z. Phys. B {\bf 94}, 469 (1994).

\bibitem{oganesyan} V. Oganesyan and I. Ussishkin, Phys. Rev. B {\bf 70}, 054503 (2004).

\bibitem{kohn} W.~Kohn, Phys. Rev. {\bf 123}, 1242 (1961).

\bibitem{fg} M.~P.~A.~Fisher and G.~Grinstein, Phys. Rev. Lett. {\bf 60}, 208 (1988).

\bibitem{jinwu} J.~Ye, Phys. Rev. B {\bf 58}, 9450 (1998).

\bibitem{bala} A.~V.~Balatsky and P.~Bourges, Phys. Rev. Lett. {\bf 82}, 5337 (1999).

\bibitem{taillefer} N.~Doiron-Leyraud, C.~Proust, D.~LeBoeuf, J.~Levallois, J.-B.~Bonnemaison,
R.~Liang, D.~A.~Bonn, W.~N.~Hardy, and L.~Taillefer, Nature {\bf 447}, 565 (2007).

\bibitem{larkin05} A. Larkin and A. Varlamov, {\em Theory of Fluctuations in Superconductors} Clarendon, Oxford (2005).

\bibitem{obra} Yu. N. Obraztsov, Fiz. Tverd. Tela {\bf 6}, 414 (1964) [Sov. Phys.—Solid State {\bf 6}, 331 (1964)]; {\bf 7}, 455 (1965).

\bibitem{sillin}  P. S. Zyryanov and V. P. Silin, Phys. Met. Metallogr. (USSR) {\bf 17}, 130 (1964).

\bibitem{streda} H.~Oji and P.~Streda, Phys. Rev. B {\bf 31}, 7291 (1985).


\bibitem{cooper} N.~R.~Cooper, B.~I.~Halperin, and I.~M.~Ruzin, Phys.
Rev. B {\bf 55}, 2344 (1997).

\bibitem{degroot} S. R. de Groot {\em The Maxwell equations: Non-relativistic and relativistic derivations from electron theory},  North-Holland Pub. Co., Amsterdam (1969).

\bibitem{demian} M.~Demianski,
{\em Relativistic Astrophysics}, Pergamon Press, Oxford (1985), Section 2.5.

\bibitem{ll} L.~D.~Landau and E.~M.~Lifshitz, {\em Fluid Mechanics\/},
Butterworth-Heinemann, Oxford (1987), Section 127.

\bibitem{sonstarinets} D.~T.~Son and A.~O.~Starinets, JHEP {\bf 0603}, 052 (2006).

\bibitem{llem}  L.~D.~Landau and E.~M.~Lifshitz, {\em The
Classical Theory of Fields\/}, Butterworth-Heinemann, Oxford (1987), Section 33.

\bibitem{halphoh} B.~I.~Halperin and P.~C.~Hohenberg, Rev. Mod. Phys.
{\bf 49}, 435 (1977).

\bibitem{km} L.~P.~Kadanoff and P.~C.~Martin, Annals of
Physics, {\bf 24}, 419 (1963).

\bibitem{Gravitation} C.~W.~Misner, K.~S.~Thorne, and J.~A.~Wheeler,
{\em Gravitation\/}, W. H. Freeman, San Francisco, (1973), Chapter 22.

\bibitem{forster} D.~Forster, {\em Hydrodynamic Fluctuations, Broken Symmetry, and Correlation
Functions\/},
Benjamin-Cummings, Reading (1975), Chapter 5.

\bibitem{giam} T.~Giamarchi, Phys. Rev. B {\bf 44}, 2905 (1991).

\bibitem{vbs}  M.~Vojta, C.~Buragohain, and S.~Sachdev, Phys. Rev. B {\bf 61}, 15152 (2000).


\bibitem{Maldacena:1997re}
  J.~M.~Maldacena,
  Adv.\ Theor.\ Math.\ Phys.\  {\bf 2}, 231 (1998)
  [Int.\ J.\ Theor.\ Phys.\  {\bf 38}, 1113 (1999)]
  [arXiv:hep-th/9711200].

\bibitem{Witten:1998zw}
  E.~Witten,
  Adv.\ Theor.\ Math.\ Phys.\  {\bf 2}, 505 (1998)
  [arXiv:hep-th/9803131].


\bibitem{pethick} G.~Baym and C.~Pethick, {\em Landau Fermi-Liquid Theory: Concepts and Applications\/},
Wiley-Interscience, New York (1991).

\bibitem{vojta} M.~Vojta, Y.~Zhang, and S.~Sachdev, Int. J. Mod. Phys. B
{\bf 14}, 3719 (2000); T. Senthil (unpublished).


\bibitem{hawking} S.~W.~Hawking and G.~F.~R.~Ellis, {\em The large scale structure
of space-time}, Cambridge University Press, Cambridge (1973).

\bibitem{moreno} J.~Moreno and P.~Coleman, arXiv:cond-mat/9603079.

\end{thebibliography}
\end{document}